\documentclass[12pt,letterpaper,english,aps,letter,prl,superscriptaddress,nofootinbib]{revtex4}
\pdfoutput=1
\usepackage[T1]{fontenc}
\usepackage[latin1]{inputenc}
\usepackage{verbatim}
\usepackage{amsmath}
\usepackage{amssymb}
\usepackage{graphicx}

\makeatletter

\pdfpageheight\paperheight
\pdfpagewidth\paperwidth

\@ifundefined{textcolor}{}
{%
 \definecolor{BLACK}{gray}{0}
 \definecolor{WHITE}{gray}{1}
 \definecolor{RED}{rgb}{1,0,0}
 \definecolor{GREEN}{rgb}{0,1,0}
 \definecolor{BLUE}{rgb}{0,0,1}
 \definecolor{CYAN}{cmyk}{1,0,0,0}
 \definecolor{MAGENTA}{cmyk}{0,1,0,0}
 \definecolor{YELLOW}{cmyk}{0,0,1,0}
}

\usepackage{ae,aecompl}

\renewcommand{\citet}[1]{\cite{#1}}

\makeatother

\usepackage{babel}
\begin{document}
\setlength{\abovedisplayskip}{0.4ex}\setlength{\belowdisplayskip}{0.4ex}
\setlength{\abovedisplayshortskip}{0.15ex}\setlength{\belowdisplayshortskip}{0.15ex}

\title{A reversible mesoscopic model of diffusion in liquids: from giant
fluctuations to Fick's law}

\author{Aleksandar Donev}

\email{donev@courant.nyu.edu}

\selectlanguage{english}%

\affiliation{Courant Institute of Mathematical Sciences, New York University,
New York, NY 10012}

\author{Thomas G. Fai}

\affiliation{Courant Institute of Mathematical Sciences, New York University,
New York, NY 10012}

\author{Eric Vanden-Eijnden}

\email{eve2@courant.nyu.edu}

\selectlanguage{english}%

\affiliation{Courant Institute of Mathematical Sciences, New York University,
New York, NY 10012}
\begin{abstract}
We study diffusive mixing in the presence of thermal fluctuations
under the assumption of large Schmidt number. In this regime we obtain
a limiting equation that contains a diffusive stochastic drift term
with diffusion coefficient obeying a Stokes-Einstein relation, in
addition to the expected advection by a random velocity. The overdamped
limit correctly reproduces both the enhanced diffusion in the ensemble-averaged
mean and the long-range correlated giant fluctuations in individual
realizations of the mixing process, and is amenable to efficient numerical
solution. Through a combination of Eulerian and Lagrangian numerical
methods we demonstrate that diffusion in liquids is not most fundamentally
described by Fick's irreversible law; rather, diffusion is better
modeled as reversible random advection by thermal velocity fluctuations.
We find that the diffusion coefficient is effectively renormalized
to a value that depends on the scale of observation. Our work reveals
somewhat unexpected connections between flows at small scales, dominated
by thermal fluctuations, and flows at large scales, dominated by turbulent
fluctuations.
\end{abstract}
\maketitle
\global\long\def\V#1{\boldsymbol{#1}}
\global\long\def\M#1{\boldsymbol{#1}}
\global\long\def\Set#1{\mathbb{#1}}

\global\long\def\D#1{\Delta#1}
\global\long\def\d#1{\delta#1}

\global\long\def\norm#1{\left\Vert #1\right\Vert }
\global\long\def\abs#1{\left|#1\right|}

\global\long\def\grad{\boldsymbol{\nabla}}
\global\long\def\av#1{\langle#1\rangle}

\global\long\def\T#1{\Tilde{#1}}
\global\long\def\Sc{\text{Sc}}

\section{Introduction}

Diffusion is one of the most ubiquitous transport processes. It is,
arguably, the simplest dissipative mechanism. Fick's law of diffusion
is ``derived'' in most elementary textbooks, and relates diffusive
fluxes to the gradient of chemical potentials via a diffusion coefficient
that is typically thought of as an independent material property.
There are several well-known hints that diffusion in liquids is, in
fact, a rather subtle process. A first hint is that the Stokes-Einstein
(SE) prediction for the diffusion coefficient is in surprisingly reasonable
agreement with measurements even in cases where it should not apply
at all, such as molecular diffusion. The fact that the SE prediction
involves the viscosity of the fluid, a seemingly independent transport
property, hints at the connection between momentum transport and diffusion.
A second hint is the fact that nonequilibrium diffusive mixing is
known to be accompanied by ``giant'' long-range correlated thermal
fluctuations \citet{LongRangeCorrelations_MD,GiantFluctuations_Universal,FluctHydroNonEq_Book}.
The enhancement of large-scale (small wavenumber) concentration fluctuations
during free diffusive mixing has been measured using light scattering
and shadowgraphy techniques \citet{GiantFluctuations_Nature,GiantFluctuations_Universal,GiantFluctuations_Cannell,FractalDiffusion_Microgravity}.

In either gases, liquids or solids, one can, at least in principle,
coarse-grain Hamiltonian dynamics for the atoms (at the classical
level) to obtain a model of diffusive mass transport at hydrodynamic
scales. The actual coarse-graining procedure is, however, greatly
simplified by first coarse-graining the microscopic dynamics to a
simpler stochastic description. In the case of gases, mass transport
can be modeled effectively using kinetic theory with cross-sections
obtained from the underlying molecular interactions. In solids, atoms
remain trapped around the crystal lattice sites for long periods of
time and infrequently hop from site to site, so that diffusive transport
can be modeled effectively at the microscopic level as a Markov Chain
with transition rates that can be obtained from the molecular interactions
using transition state theory. In both of these cases the picture
that emerges is that of independent Brownian walkers performing uncorrelated
random walks in continuum (gases) or on a lattice (solids). By contrast,
in liquids the physical picture is rather different and must account
for hydrodynamic correlations among the diffusing particles. In a
liquid, molecules become trapped (caged) over long periods of time,
as they collide frequently with their neighbors. Therefore, momentum
and energy are exchanged (diffuse) much faster than the molecules
themselves can escape their cage. The main mechanism by which molecules
diffuse is the motion of the whole cage when a large-scale velocity
fluctuation (coordinated motion of parcels of fluid) moves a group
of molecules and shifts and rearranges the cage. 

It is now well-understood that diffusion in liquids is strongly affected
by advection by thermal velocity fluctuations \citet{DiffusionRenormalization_I,ExtraDiffusion_Vailati,DiffusionRenormalization,Nanopore_Fluctuations}.
The fact that thermal fluctuations exhibit long-ranged correlations
in nonequilibrium settings has long been appreciated in statistical
mechanics and nonequilibrium thermodynamics circles \citet{LongRangeCorrelations_MD,FluctHydroNonEq_Book}.
The overarching importance of nonequilibrium fluctuations to transport
in fluids has not, however, been widely appreciated. The microgravity
experiments described in Ref. \citet{FractalDiffusion_Microgravity}
show fluctuations of the order of a fraction of a percent at millimeter
scales. These results are a striking demonstration that thermal fluctuations
are important not just at microscopic and mesoscopic scales, but also
at \emph{macroscopic} scales. Theoretical studies and computer simulations
have verified that the advection by the thermally fluctuating fluid
velocity leads to an \emph{enhancement} or \emph{renormalization }of
the diffusion coefficient that depends on the viscosity of the fluid,
and, importantly, on the dimensionality and imposed boundary conditions
(in particular, system size) \citet{ExtraDiffusion_Vailati,Nanopore_Fluctuations,DiffusionRenormalization}.
At the mathematical level, the diffusion enhancement is closely-related
to the eddy diffusivity that arises in turbulent flows as mass is
advected by the chaotic fluid velocity \citet{TurbulenceClosures_Majda}.
When modeling molecular diffusion, most theories are based on some
form of mode-mode coupling, which is essentially a perturbative analysis
in the strength of the thermal fluctuations \citet{DiffusionRenormalization_I,DiffusionRenormalization_II},
starting from linearized fluctuating hydrodynamics. When modeling
diffusion of particles suspended in a (complex) fluid, it is typically
assumed that the immersed particle is either very large, much more
massive, or both, compared to the fluid molecules \citet{DiffusionRenormalization_III}.
In this work, for the first time, we bypass these types of approximations
and develop a fully nonlinear fluctuating hydrodynamic description
of diffusion that applies not only to a dilute suspension of large
particles suspended in a simple liquid but also to molecular (self)
diffusion of tagged molecules.

Here we formulate a simple model for diffusion in the presence of
thermal velocity fluctuations and use it to make a precise assessment
of the contribution of fluctuations to diffusive transport. In our
model the momentum exchange is modeled using a continuum fluctuating
hydrodynamic formalism, and represents the background fluctuating
momentum (velocity) bath with which the diffusing particles (tracers)
interact. We demonstrate that this simple model mimics all of the
crucial features of realistic liquids, while also being tractable
analytically and numerically, and showing rich physical behavior.
Along the way we will construct a multiscale numerical method that
can efficiently handle the practically-relevant case of very large
Schmidt number. In our model we assume that the diffusing particles
follow, on average, the locally-averaged (thermally fluctuating) velocity
of the fluid. We further assume the existence of a large separation
of time scales between the fast dynamics of the velocity (vorticity)
fluctuations and the diffusive dynamics, i.e., we assume that the
Schmidt number $\text{Sc}$ (ratio of the diffusion coefficients of
momentum and mass) is very large. This is known to be true in most
liquids due to the effective caging of molecules in densely-packed
liquid microstructures. In the limit $\text{Sc}\rightarrow\infty$
we can eliminate the fluid velocity using adiabatic mode elimination
\citet{AdiabaticElimination_1,ModeElimination_Papanicolaou} and obtain
explicit results without resulting to perturbation analysis, as have
previous studies.

Through a mix of theoretical and numerical studies, in this paper
we demonstrate that over a broad range of length scales diffusion
is better described as a reversible stochastic process, rather than
an irreversible deterministic process as in Fick's law. This has the
consequence that the diffusion coefficient (effective dissipation
rate) is not a material constant, but rather depends on the scale
of observation. Our work uncovers crucial distinctions between the
Fickian diffusion of uncorrelated Brownian walkers and the non-Fickian
collective diffusion of hydrodynamically-correlated tracers. We consider
non-interacting particles, and make an important first step toward
consistently including the effects of hydrodynamic correlations in
fluctuating Dynamic Density Functional Theory (DDFT) \citet{DDFT_Diffusion,DDFT_Pep,SPDE_Diffusion_DDFT,SPDE_Diffusion_Dean,SPDE_Diffusion_Formal}
for suspensions of interacting colloids. Importantly, our work demonstrates
that Fick's law for the average concentration is not affected by the
presence of hydrodynamic interactions, and the non-local diffusion
terms proposed in prior works \citet{DDFT_Lowen,DDFT_Pavliotis_PRL}
do not actually appear. Some of our predictions could be used as a
guide to design new Fluorescence Recovery After Photo-bleaching (FRAP)
experiments to study the collective dynamics of tracer particles in
liquids.

We begin by outlining our starting fluctuating hydrodynamics model
for diffusion in the presence of thermal velocity fluctuations. These
equations show extreme numerical stiffness at large Schmidt numbers
and in Section \ref{sec:infSc} and Appendix \ref{sec:ModeElimination}
we obtain the limit of these equations as the Schmidt numbers becomes
infinite. In Appendix \ref{sub:Algorithm} we design Eulerian and
Lagrangian numerical algorithms to solve the resulting advection-diffusion
equation. In Section \ref{sec:Irreversibility} we numerically and
analytically study the difference between (dissipative and irreversible)
classical Fickian diffusion and (conservative and reversible) diffusion
by thermal velocity fluctuations, with particular focus on the appearance
of giant fluctuations in the latter. In Section \ref{sec:SpatialCG}
we discuss spatial coarse-graining of the limiting advection-diffusion
equation, and summarize the emerging paradigm for how to accurately
model diffusion in liquids over a broad range of scales. Finally,
we offer some conclusions and a discussion of open challenges in Section
\ref{sec:Conclusions}.

\subsection{Advection-Diffusion Model}

We consider the diffusion of passive tracer particles as they are
advected by thermal velocity fluctuations. Examples include the diffusion
of fluorescently-labeled macromolecules in solution, nano-colloidal
particles in a nanofluid, and the self-diffusion of the molecules
comprising a simple fluid. The hydrodynamic fluctuations of the fluid
velocity $\V v\left(\V r,t\right)$ will be modeled via the linearized
incompressible fluctuating Navier-Stokes equation in $d$ dimensions,
$\grad\cdot\V v=0$, and
\begin{equation}
\rho\partial_{t}\V v+\grad\pi=\eta\grad^{2}\V v+\sqrt{2\eta k_{B}T}\,\grad\cdot\M{\mathcal{W}},\label{eq:v_eq}
\end{equation}
where $\rho$ is the fluid density, $\eta$ the viscosity, and $T$
the temperature, all assumed to be constant throughout the domain,
and $\pi\left(\V r,t\right)$ is the mechanical pressure. Here the
stochastic momentum flux is modeled via a a white-noise symmetric
tensor field $\M{\mathcal{W}}\left(\V r,t\right)$ with covariance
chosen to obey a fluctuation-dissipation principle \citet{FluctHydroNonEq_Book,OttingerBook},
\[
\av{\mathcal{W}_{ij}(\V r,t)\mathcal{W}_{kl}(\V r^{\prime},t^{\prime})}.=\left(\delta_{ik}\delta_{jl}+\delta_{il}\delta_{jk}\right)\delta(t-t^{\prime})\delta(\V r-\V r^{\prime}).
\]
Note that because the noise is additive in (\ref{eq:v_eq}) there
is no difference between an Ito and a Stratonovich interpretation
of the stochastic term.

The details of the coupling between the fluid and the passive tracer
are complicated at the microscopic level \citet{StokesEinstein_BCs}
and some approximations are required to model the motion of the tracer.
The principal effect of advection by the thermal velocity fluctuations
can be captured by assuming that the position of a tracer $\V q\left(t\right)$
follows a spatially smooth fluctuating velocity field $\V u\left(\V r,t\right)$,
\begin{equation}
d\V q/dt=\V u\left(\V q,t\right)+\sqrt{2\chi_{0}}\,\M{\mathcal{W}}_{\V q}\left(t\right),\label{eq:Lagrangian_eq}
\end{equation}
where $\M{\mathcal{W}}_{\V q}$ denotes a collection of $d$ independent
white-noise processes. Here $\chi_{0}$ is a \emph{bare} diffusion
coefficient that can be thought of as representing a ``random slip''
relative to the local fluid velocity coming from the under-resolved
microscopic dynamics, \emph{uncorrelated} among distinct tracers.
In what follows it will be crucial that $\V u$ be divergence free,
$\grad\cdot\V u=0$. In this work we assume that the velocity felt
by the tracer, 
\begin{equation}
\V u\left(\V r,t\right)=\int\M{\sigma}\left(\V r,\V r^{\prime}\right)\V v\left(\V r^{\prime},t\right)d\V r^{\prime}\equiv\M{\sigma}\star\V v,\label{eq:smoothing_u}
\end{equation}
is obtained by convolving %
\footnote{In translationally-invariant systems (e.g., infinite or periodic systems)
one can use standard convolution $\M{\sigma}\left(\V r,\V r^{\prime}\right)\equiv\M{\sigma}\left(\V r-\V r^{\prime}\right)$
and would typically take an isotropic kernel $\M{\sigma}\left(\V r-\V r^{\prime}\right)=K_{\sigma}\left(\norm{\V r-\V r^{\prime}}\right)$,
where $K_{\sigma}\left(r\right)$ is a symmetric ``bell-shaped''
function with support of length $2\sigma$. When nontrivial boundary
conditions are present (i.e., for confined systems), it is important
to consider a more general filtering operation (mollification) that
takes into account the boundary conditions. For certain types of boundary
conditions the construction of $\M{\sigma}\left(\V r,\V r^{\prime}\right)$
is nontrivial and ought be done on a case-by-case basis \citet{alphaNS_Titi}.%
} the fluid velocity with a smoothing kernel $\M{\sigma}$ that filters
out features at scales below a molecular cutoff scale $\sigma$. For
example, in the $\alpha$-Navier-Stokes equations \citet{alphaNS_Titi}
the smoothing is chosen to be an inverse Helmholtz operator, $\V v=\M u-\sigma^{2}\grad^{2}\V u$,
with boundary conditions chosen such that $\V u$ is divergence free
within the domain of interest \citet{alphaNS_BCs}. With periodic
boundary conditions, in Fourier space, the inverse Helmholtz operator
filter is $\hat{\M{\sigma}}_{\V k}=\left(1+\sigma^{2}k^{2}\right)^{-1}\M I$.

It is important to point out that the smoothing or regularization
of the fluctuating velocity field (\ref{eq:smoothing_u}) is necessary.
Otherwise, the diffusion coefficient of the tracer particle will diverge
leading to an ``ultraviolet catastrophe'' familiar in renormalization
theories. In the literature a phenomenological cutoff at large wavenumbers
is imposed \citet{DiffusionRenormalization_I,DiffusionRenormalization_II,ExtraDiffusion_Vailati}.
We implement this regularization here by applying a smooth filter
to $\V v$ to generate a velocity $\V u$ with which we can advect
tracers. Alternatively, one can filter the white-noise forcing in
the velocity equation and replace $\M{\mathcal{W}}$ by $\M{\sigma}\star\M{\mathcal{W}}$
in (\ref{eq:v_eq}). In the end, as far as the passive tracers are
concerned, the only thing that matters is the spatio-temporal spectrum
of the advective velocity $\V u\left(\V r,t\right)$.

Let us assume that there are $N$ tracers and define the concentration
or density of tracers $c\left(\V r,t\right)=\sum_{i=1}^{N}\delta\left(\V q_{i}\left(t\right)-\V r\right)$.
The Lagrangian description (\ref{eq:Lagrangian_eq}) of the dynamics
of the individual tracers formally corresponds to an Eulerian description
for the evolution of the concentration or number density of tracers
$c\left(\V r,t\right)$ via a fluctuating advection-diffusion Ito
equation \citet{SPDE_Diffusion_Dean,SPDE_Diffusion_Formal},
\begin{equation}
\partial_{t}c=-\V u\cdot\grad c+\chi_{0}\grad^{2}c+\grad\cdot\left(\sqrt{2\chi_{0}c}\,\M{\mathcal{W}}_{c}\right),\label{eq:c_eq_original}
\end{equation}
where $\M{\mathcal{W}}_{c}\left(\V r,t\right)$ denotes a white-noise
vector field. It is important to point out that this equation is simply
a formal rewriting of the equations of motion (\ref{eq:Lagrangian_eq})
for the $N$ tracers and as such contains no new physical content.
However, it can be argued that (\ref{eq:c_eq_original}) also describes
the dynamics of a spatially coarse-grained \emph{smooth }concentration
field when the density of tracers varies on a length scale much larger
than the typical tracer distance \citet{SPDE_Diffusion_DDFT,DiscreteDiffusion_Espanol}.
While a precise mathematical derivation of (\ref{eq:c_eq_original})
from (\ref{eq:Lagrangian_eq}) is unavailable at present except in
the case of no bare diffusion, $\chi_{0}=0$, we believe it is a very
plausible fluctuating hydrodynamic model of self-diffusion or diffusion
of dilute passive tracers in liquids.

\begin{figure*}[!t]
\begin{centering}
\includegraphics[width=0.4\textwidth]{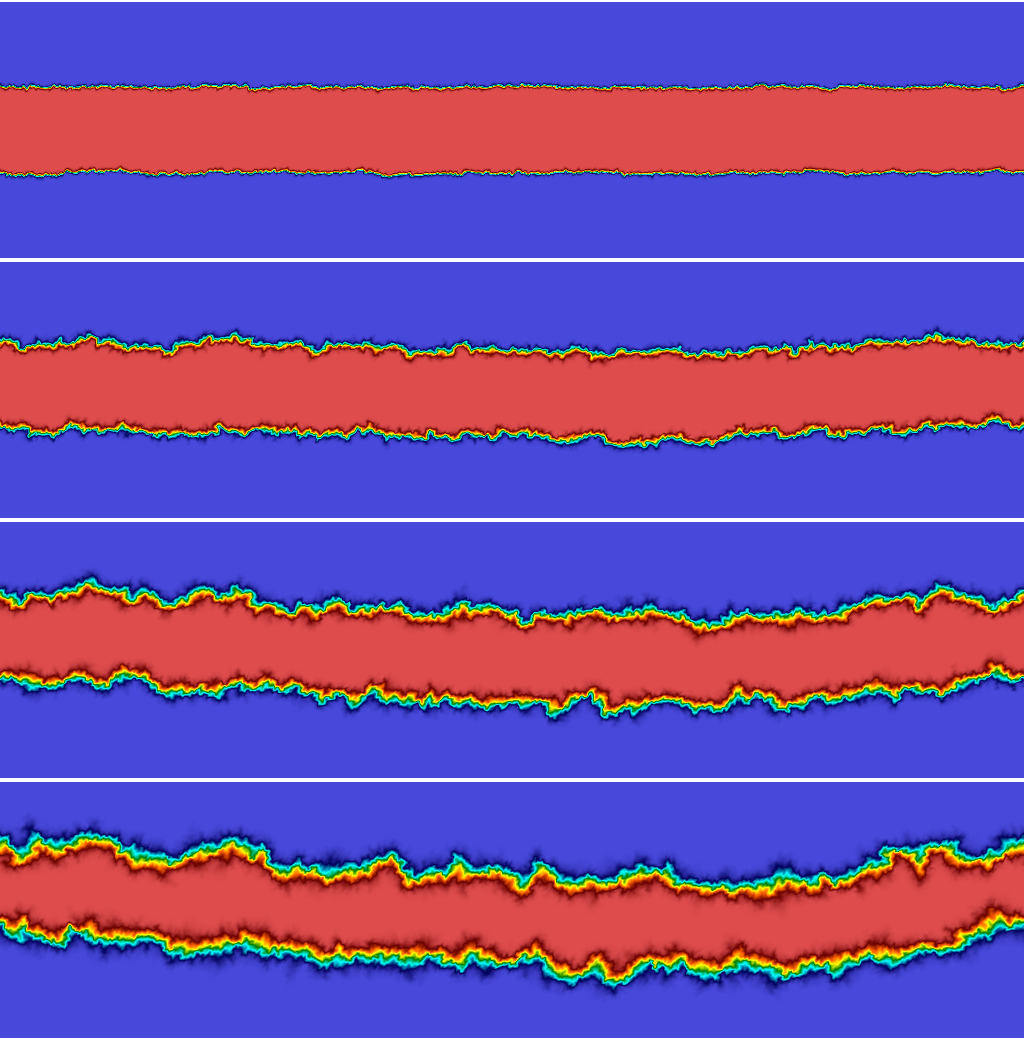}\hspace{0.5cm}\includegraphics[width=0.4\textwidth]{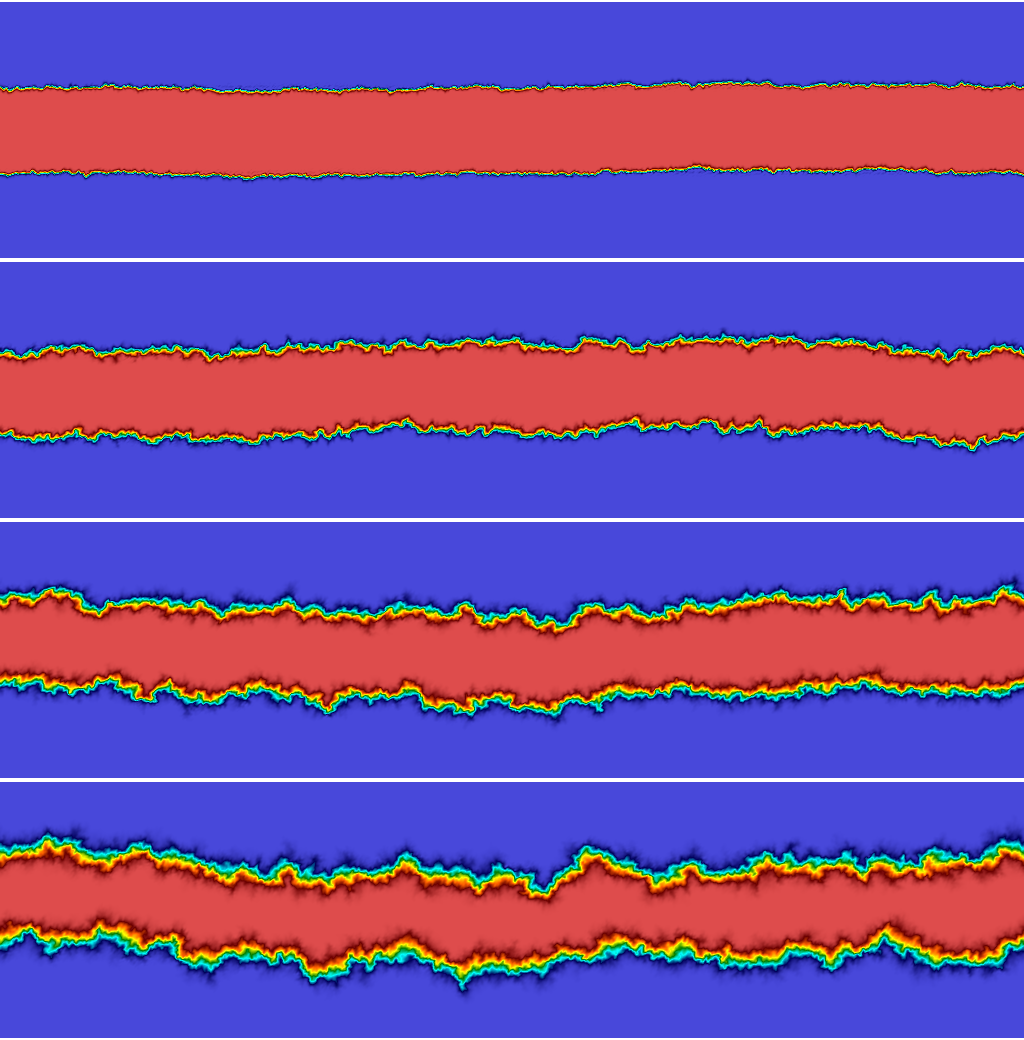}
\par\end{centering}

\caption{\label{fig:DiffusiveInterface}Snapshots of concentration showing
the development of a \emph{rough} diffusive interface between two
miscible fluids, starting from concentration being unity in a horizontal
(red) stripe occupying one third of the periodic domain, and zero
elsewhere. We show four snapshots in time (evolving from top to bottom).
The effective Schmidt number is $\Sc\approx1.5\cdot10^{3}$ and the
bare diffusion is such that $\chi/\chi_{0}\approx5$. (\emph{Left
panel}) A single instance of the resolved dynamics (\ref{eq:v_eq},\ref{eq:c_eq_original}).
(\emph{Right panel}) An independent instance of the limiting dynamics
(\ref{eq:limiting_Strato}), or, equivalently, (\ref{eq:limiting_Ito}),
obtained at a very small fraction ($10^{-3}\sim\Sc^{-1}$, see Section
\ref{sub:EulerianAlgorithm}) of the computational cost of the simulation
shown in the left panel.}
\end{figure*}

As an illustration of the importance of thermal fluctuations in diffusive
transport we use recently-developed finite-volume numerical methods
\citet{LLNS_Staggered,DFDB} for solving (\ref{eq:v_eq},\ref{eq:c_eq_original})
to model the diffusive mixing between two initially phase-separated
fluids in two dimensions with periodic boundary conditions. In the
left panel of Fig. \ref{fig:DiffusiveInterface} we show snapshots
of the concentration field at several points in time. As seen in the
figure, the interface between the fluids develops large-scale roughness
(giant fluctuations) instead of remaining flat as in simple diffusion
\citet{FractalDiffusion_Microgravity}. This roughening is accompanied
by a slow spreading of the initially-sharp interface, similarly to
what would be observed in deterministic diffusion. Molecular dynamics
simulations have confirmed that fluctuating hydrodynamics accurately
models the diffusive mixing process down to essentially molecular
scales \citet{LowMachExplicit}.

\section{\label{sec:infSc}The Limit of Large Schmidt Number}

In liquids, diffusion of mass is much slower than diffusion of momentum,
i.e., the Schmidt number is very large. More precisely, there is a
large separation of time scales between the fast dynamics of the velocity
fluctuations and the slow evolution of the concentration. This separation
of time scales, to be verified \emph{a posteriori,} can be used to
perform a formal adiabatic mode-elimination procedure of the fast
velocity degrees of freedom \citet{Averaging_Khasminskii,Averaging_Kurtz,ModeElimination_Papanicolaou,AveragingHomogenization}
in (\ref{eq:v_eq},\ref{eq:c_eq_original}). The mode-elimination
procedure, which is detailed in Appendix \ref{sec:ModeElimination},
gives a \emph{limiting }stochastic advection-diffusion equation for
the overdamped dynamics of the concentration, 
\begin{equation}
\partial_{t}c=-\V w\odot\grad c+\chi_{0}\grad^{2}c+\grad\cdot\left(\sqrt{2\chi_{0}c}\,\M{\mathcal{W}}_{c}\right),\label{eq:limiting_Strato}
\end{equation}
where $\odot$ denotes a Stratonovich dot product, and the advection
velocity $\V w\left(\V r,t\right)$ is white in time, with covariance
proportional to a Green-Kubo integral of the velocity auto-correlation
function, 
\begin{align}
\av{\V w\left(\V r,t\right)\otimes\V w\left(\V r^{\prime},t^{\prime}\right)} & =\M{\mathcal{R}}\left(\V r,\V r^{\prime}\right)\delta\left(t-t^{\prime}\right),\label{eq:C_w}
\end{align}
\begin{equation}
\M{\mathcal{R}}\left(\V r,\V r^{\prime}\right)=2\int_{0}^{\infty}\av{\V u\left(\V r,t\right)\otimes\V u\left(\V r^{\prime},t+t^{\prime}\right)}dt^{\prime}.\label{eq:R_r}
\end{equation}
The term $\V w\odot\grad c$ is reminiscent of the random advection
in the Kraichnan model of turbulent transport of a passive tracer
\citet{Kraichnan_SelfSimilar,Kraichnan_SelfSimilarPRL} (see section
4.1 in \citet{TurbulenceClosures_Majda}).

To be more precise, let us perform a spectral decomposition of the
covariance $\M{\mathcal{R}}$ in some (infinite dimensional) set of
(non-normalized) eigenfunctions $\V{\phi}_{k}$, 
\begin{equation}
\M{\mathcal{R}}\left(\V r,\V r^{\prime}\right)=\sum_{k}\V{\phi}_{k}\left(\V r\right)\otimes\V{\phi}_{k}\left(\V r^{\prime}\right).\label{eq:R_spectral}
\end{equation}
The notation $\V w\odot\grad c$ is short-hand for $\sum_{k}\left(\V{\phi}_{k}\cdot\grad c\right)\circ d\mathcal{B}_{k}/dt$,
where $\mathcal{B}_{k}\left(t\right)$ are independent Brownian motions
(Wiener processes). In our numerical simulations, we use the Stratonovich
form of the equations and apply an Euler-Heun (midpoint predictor-corrector)
temporal integrator to (\ref{eq:limiting_Strato}).

For calculating ensemble averages, the Ito form of the equation is
more useful. In the Ito interpretation, as derived in Appendix \ref{sec:ModeElimination},
an unexpected ``thermal'' or ``Ito'' drift appears in the limiting
equation and takes the form of an \emph{enhanced} diffusion,
\begin{align}
\partial_{t}c & =-\V w\cdot\grad c+\grad\cdot\left[\M{\chi}\left(\V r\right)\grad c\right]+\chi_{0}\grad^{2}c+\grad\cdot\left(\sqrt{2\chi_{0}c}\,\M{\mathcal{W}}_{c}\right)\label{eq:limiting_Ito}
\end{align}
where the enhancement of the diffusion coefficient is given by the
integral of the velocity autocorrelation function, 
\begin{equation}
\M{\chi}\left(\V r\right)=\tfrac{1}{2}\M{\mathcal{R}}\left(\V r,\V r\right)=\int_{0}^{\infty}\av{\V u\left(\V r,t\right)\otimes\V u\left(\V r,t+t^{\prime}\right)}dt^{\prime}.\label{eq:chi_r_general}
\end{equation}
Here we have made use of the fact that $\grad\cdot\V w=0$, and $\V w\cdot\grad c$
is shorthand notation for $\sum_{k}\left(\V{\phi}_{k}\cdot\grad c\right)d\mathcal{B}_{k}/dt$.
The Ito equation (\ref{eq:limiting_Ito}) is the key result of the
mode elimination procedure. The last two terms in this equation are
deterministic diffusive terms, while the advective term $-\V w\cdot\grad c$
is a (multiplicative) stochastic noise term that vanishes in the mean
as a consequence of the Ito interpretation. That is, the ensemble
average of the concentration obeys Fick's law,
\begin{equation}
\partial_{t}\av c=\grad\cdot\left(\M{\chi}_{\text{eff}}\grad\av c\right)=\grad\cdot\left[\left(\chi_{0}+\M{\chi}\right)\grad\av c\right],\label{eq:dc_dt_mean}
\end{equation}
which is a well-known result (c.f. (255) in Ref. \citet{ModeElimination_MTV})
that can be justified rigorously using stochastic homogenization theory
\citet{ModeElimination_MTV}. In the absence of bare diffusion, $\chi_{0}=0$,
the same equation (\ref{eq:dc_dt_mean}) holds for \emph{all} moments
of $c$.

We note that (\ref{eq:limiting_Ito}) looks like an expected result
but this is deceptive. While indeed the Green-Kubo expression for
the diffusion coefficient of a tracer particle is well-known to be
(\ref{eq:chi_r_general}), what (\ref{eq:limiting_Ito}) is describing
is the \emph{collective} and not the \emph{individual} diffusion of
a tagged particle. The subtle difference between the two stems from
the importance of the hydrodynamic correlations among the trajectories
of the tracers. The ``dissipative'' term $\grad\cdot\left[\M{\chi}\left(\V r\right)\grad c\right]$
and the ``fluctuating'' term $-\V w\cdot\grad c$ are signatures
of the same physical process, advection by thermal velocity fluctuations.
This is most clearly seen in the Stratonovich form (\ref{eq:limiting_Strato})
where there is only a single stochastic term $-\V w\odot\grad c$
present. The stochastic advection term in (\ref{eq:limiting_Ito})
need to be retained to obtain the giant fluctuations seen in a \emph{particular
instance} (realization) of the diffusive mixing process. Including
the dissipative term $\grad\cdot\left[\M{\chi}\left(\V r\right)\grad c\right]$
but omitting the random advection term $-\V w\cdot\grad c$ violates
fluctuation-dissipation balance and \emph{cannot} be justified by
simply arguing that the fluctuating term has mean zero. Just like
the stochastic noise term $\grad\cdot\left(\sqrt{2\chi_{0}c}\,\M{\mathcal{W}}_{c}\right)$
corresponds to (i.e., is in fluctuation-dissipation balance with)
the dissipative term $\chi_{0}\grad^{2}c$ for a collection of uncorrelated
random walkers, the stochastic noise term $-\V w\cdot\grad c$ corresponds
to the Fickian term $\grad\cdot\left[\M{\chi}\left(\V r\right)\grad c\right]$
for a collection of hydrodynamically correlated tracers. While the
dissipation in both cases looks like simple diffusion, the important
distinction between the two types of microscopic dynamics is made
clear in the stochastic forcing term.

\subsection{The Stokes-Einstein Relation}

In the overdamped limit, the details of the evolution of the fluid
velocity do not matter, so long as there exists a unique time-reversible
equilibrium dynamics over which the average in (\ref{eq:R_r}) is
taken. If one assumes the linearized fluctuating Navier-Stokes equation
(\ref{eq:v_eq}) holds, then it is not hard to show that
\begin{equation}
\int_{0}^{\infty}\av{\V v\left(\V r,t\right)\otimes\V v\left(\V r^{\prime},t+t^{\prime}\right)}dt^{\prime}=\frac{k_{B}T}{\eta}\M G\left(\V r,\V r^{\prime}\right),\label{eq:v_cov_Stokes}
\end{equation}
where $\M G$ is the Green's function (Oseen tensor) for the steady
Stokes equation %
\footnote{For unbounded three-dimensional systems the Oseen tensor is $\V G\left(\V r^{\prime},\V r^{\prime\prime}\right)=\left(8\pi r\right)^{-1}\left(\M I+r^{-2}\V r\otimes\V r\right)$,
where $\V r=\V r^{\prime}-\V r^{\prime\prime}$.%
} with unit viscosity, $\V v=\M G\star\V f$ if $\grad\pi=\grad^{2}\V v+\V f$
subject to $\grad\cdot\V v=0$ and appropriate boundary conditions.
A sample of the Brownian increment $\sum_{k}\V{\phi}_{k}d\mathcal{B}_{k}$
can be obtained by solving a steady Stokes problem with a suitable
random forcing (fluctuating stress), and then convolving the velocity
with the filter $\M{\sigma}$. 

The diffusion enhancement (\ref{eq:chi_r_general}) can be obtained
explicitly from (\ref{eq:v_cov_Stokes}) as 
\begin{equation}
\M{\chi}\left(\V r\right)=\frac{k_{B}T}{\eta}\int\M{\sigma}\left(\V r,\V r^{\prime}\right)\M G\left(\V r^{\prime},\V r^{\prime\prime}\right)\M{\sigma}^{T}\left(\V r,\V r^{\prime\prime}\right)d\V r^{\prime}d\V r^{\prime\prime}.\label{eq:chi_r_Stokes}
\end{equation}
The relation (\ref{eq:chi_r_Stokes}) is nothing more than the Stokes-Einstein
(SE) relation $\M{\chi}\left(\V q\right)=k_{B}T\,\M{\mu}\left(\V q\right)$,
where $\M{\mu}$ is the \emph{deterministic} mobility of one of the
tracers, defined via the relation $\V u=\M{\mu}\V F$, where $\V F$
is a constant force applied on a tracer at position $\V q$ and $\V u$
is the resulting velocity of the tracer \citet{StokesEinstein}. Namely,
the applied force can easily be included in our model as an additional
force density $\V f\left(\V r\right)=\M{\sigma}^{T}\left(\V q,\V r\right)\V F$
in the steady Stokes equation \citet{SELM,StokesEinstein}, which
directly leads to the SE relation (\ref{eq:chi_r_Stokes}). This shows
that the diffusion enhancement $\M{\chi}\left(\V r\right)$ is consistent
with the SE relation, as expected, validating the model. 

To get an intuitive understanding and an estimate of the diffusion
enhancement $\M{\chi}$ we consider an infinite isotropic system and
introduce a cutoff for the fluctuations in the advective velocity
$\V w$ at \emph{both} large and small scales. The large-scale cutoff
corresponds to a finite extent of the system $L$, and the small-scale
cutoff corresponds to the filtering at the molecular scale $\sigma$.
According to (\ref{eq:v_cov_Stokes}), at intermediate scales the
Fourier spectrum of $\V w$ should match the Green's function for
Stokes flow with unit density and viscosity, $\widehat{\M G}_{\V k}=k^{-2}\left(\M I-k^{-2}\V k\otimes\V k\right)$
for wavevector $\V k$. As an example, we choose an isotropic filtering
kernel $\M{\sigma}\left(\V r,\V r^{\prime}\right)=\varsigma\left(\norm{\V r-\V r^{\prime}}\right)\M I$
giving $\M{\mathcal{R}}\left(\V r,\V r^{\prime}\right)=\M{\mathcal{R}}\left(\norm{\V r-\V r^{\prime}}\right)$,
such that the Fourier transform of (\ref{eq:chi_r_Stokes}), $\hat{\M{\mathcal{R}}}_{\V k}=2\left(k_{B}T/\eta\right)\abs{\hat{\varsigma}_{\V k}}^{2}\widehat{\M G}_{\V k}$,
has the form 
\begin{equation}
\hat{\M{\mathcal{R}}}_{\V k}=\frac{2k_{B}T}{\eta}\frac{k^{2}L^{4}}{\left(1+k^{4}L^{4}\right)\left(1+k^{2}\sigma^{2}\right)}\left(\M I-\frac{\V k\otimes\V k}{k^{2}}\right).\label{eq:chi_k_filtered}
\end{equation}
This particular form is chosen for convenience and not because of
any particular physical importance; the important thing is that at
intermediate $L^{-1}\ll k\ll\sigma^{-1}$ we have $\hat{\varsigma}_{\V k}\approx1$,
that $\abs{\hat{\varsigma}_{\V k}}\ll1$ for $k\ll L^{-1}$ and vanishes
\footnote{An alternative approach is to make the lower bound for integrals in
Fourier space be $k=2\pi/L$, mimicking a Fourier series for a cubic
box of length $L$.%
} at $\V k=\V 0$, and that $\hat{\M{\mathcal{R}}}_{\V k}$ decays
faster than $k^{-d}$ for $k\gg\sigma^{-1}$. Converting (\ref{eq:chi_k_filtered})
to real space gives an isotropic enhancement to the diffusion tensor
$\M{\chi}=\M{\mathcal{R}}(0)/2=\left(2\pi\right)^{-d}\int\left(\hat{\M{\mathcal{R}}}_{\V k}/2\right)d\V k=\chi\M I$.
Note that this Fourier integral is exactly the one that appears in
the linearized steady-state (static) approximate renormalization theory
\citet{DiffusionRenormalization_I,DiffusionRenormalization_II} when
$\nu\gg\chi_{0}$ (c.f. Eq. (12) in \citet{DiffusionRenormalization}).
This is expected since in the end all approaches lead to the familiar
SE formula for the diffusion coefficient of an individual tracer (but
recall the important distinction between the individual and collective
diffusion). Our derivation shows that the SE relation can be seen
as a formula for the eddy-diffusivity due to advection by a thermally-fluctuating
random velocity field.

Performing the integral of (\ref{eq:chi_k_filtered}) in spherical
coordinates gives an asymptotic expansion in $\sigma/L$, 
\begin{equation}
\chi=\frac{k_{B}T}{\eta}\begin{cases}
\left(4\pi\right)^{-1}\ln\frac{L}{\sigma} & \mbox{if }d=2\\
\left(6\pi\sigma\right)^{-1}\left(1-\frac{\sqrt{2}}{2}\frac{\sigma}{L}\right) & \mbox{if }d=3.
\end{cases}\label{eq:chi_SE}
\end{equation}
Note that in three dimensions the coefficients $\sqrt{2}/2$ and the
$6\pi$ in the denominator in (\ref{eq:chi_SE}) depend on the exact
form of the spectrum $\hat{\M{\mathcal{\chi}}}_{\V k}$, but the coefficient
of $4\pi$ in two dimensions does \emph{not} depend on the details
of the spectrum at small and large $k$. For an isotropic Gaussian
filter $\M{\sigma}$ with standard deviation $\sigma$, as employed
in our Lagrangian numerical algorithm, for a periodic domain of length
$L$ the diffusion enhancement has a form similar to (\ref{eq:chi_SE}),
$\chi=k_{B}T\left(4\pi\eta\right)^{-1}\ln\left[L/\left(\alpha\sigma\right)\right]$
in two dimensions, where we numerically estimate the coefficient $\alpha\approx5.5$.
When $L\gg\sigma$, in three dimensions (\ref{eq:chi_SE}) gives the
Stokes-Einstein prediction $\chi\approx\chi_{SE}=k_{B}T/\left(6\pi\eta\sigma\right)$
for the diffusion coefficient of a slowly-diffusing no-slip rigid
sphere of radius $\sigma$. In two dimensions, the effective diffusion
coefficient grows logarithmically with system size, in agreement with
the Einstein relation and the Stokes paradox for the mobility of a
disk of radius $\sigma$. This system-size dependence of the effective
diffusion coefficient was quantitatively verified using steady-state
particle simulations in Refs. \citet{DiffusionRenormalization_PRL,DiffusionRenormalization}.

It is important to note that (\ref{eq:chi_r_Stokes}) continues to
hold in bounded domains also, and can be used to obtain expressions
for the position-dependent tensor diffusion coefficient $\M{\chi}\left(\V r\right)$
of a tracer particle in confined domains such as nano-channels \citet{Nanopore_Fluctuations,Nanopore_FluctuationsPRE}.
Similarly, (\ref{eq:limiting_Ito}) can describe the collective diffusion
of many tracers through a nano-channel. In confined geometry, the
Green's function $\M G$ can be expressed as an infinite series $\M G\left(\V r,\V r^{\prime}\right)=\sum_{k}\lambda_{k}^{-1}\V{\varphi}_{k}\left(\V r\right)\otimes\V{\varphi}_{k}\left(\V r^{\prime}\right)$,
where $\V{\varphi}_{k}$ are a set of orthonormal eigenfunctions of
the Stokes problem (with the appropriate boundary conditions), and
$\lambda_{k}$ are the associated eigenvalues. Note that the eigenfunctions
used to factorize the covariance of $\V w$ in (\ref{eq:R_spectral})
can be written as $\V{\phi}_{k}=\sqrt{\lambda_{k}}\,\M{\sigma}\star\V{\varphi}_{k}$
for the case when $\V v$ follows the fluctuating Stokes equation.

Having obtained and analyzed the limiting dynamics, we are in a position
to ascertain the validity of the initial assumption of large separation
of time scales between concentration and momentum diffusion. Specifically,
the limiting equations (\ref{eq:limiting_Strato},\ref{eq:limiting_Ito})
{[}similarly, (\ref{eq:Lagrangian_Strato},\ref{eq:Lagrangian_Ito}){]}
are good approximations to (\ref{eq:c_eq_original}) {[}correspondingly,
(\ref{eq:Lagrangian_eq}){]} if the effective Schmidt number $\text{Sc}=\nu/\chi_{\text{eff}}=\nu/\left(\chi_{0}+\chi\right)\gg1$,
where $\nu=\eta/\rho$ is the kinematic viscosity (momentum diffusion
coefficient). This is indeed the case in practice for both simple
liquids and especially for macromolecular solutions.

\subsection{Relation to Brownian Dynamics}

In the Lagrangian description, the overdamped limit of (\ref{eq:Lagrangian_eq})
in the Stratonovich interpretation can be shown to be
\begin{equation}
d\V q=\sum_{k}\V{\phi}_{k}\left(\V q\right)\circ d\mathcal{B}_{k}+\sqrt{2\chi_{0}}\, d\M{\mathcal{B}}_{\V q}.\label{eq:Lagrangian_Strato}
\end{equation}
This derivation is not presented here but is rather standard and follows
the procedure outlined in Appendix \ref{sec:ModeElimination}, albeit
greatly simplified by the finite-dimensional character of the limiting
equation. The second stochastic term on the right hand side of (\ref{eq:Lagrangian_Strato})
uses an independent Brownian motion $\M{\mathcal{B}}_{\V q}(t)$ for
each Brownian walker (tracer). The first stochastic forcing term uses
a \emph{single }realization of the random field $\sum_{k}\V{\phi}_{k}\circ d\mathcal{B}_{k}$
for \emph{all} of the walkers, and therefore induces correlations
between the trajectories of the tracers. In the Ito interpretation
the Lagrangian overdamped dynamics takes the form
\begin{equation}
d\V q=\sum_{k}\V{\phi}_{k}\left(\V q\right)d\mathcal{B}_{k}+\left[\partial_{\V q}\cdot\M{\chi}\left(\V q\right)\right]dt+\sqrt{2\chi_{0}}\, d\M{\mathcal{B}}_{\V q}.\label{eq:Lagrangian_Ito}
\end{equation}
For translationally-invariant systems the thermal or stochasic drift
term vanishes because $\M{\chi}$ is independent of the position of
tracer, $\partial_{\V q}\cdot\M{\chi}=\V 0$. In the more general
case, it can be shown from (\ref{eq:C_w}) that $\partial_{\V q}\cdot\M{\chi}\left(\V q\right)=\sum_{k}\V{\phi}_{k}\left(\V q\right)\cdot\grad\V{\phi}_{k}\left(\V q\right)$.

Our overdamped Lagrangian equations (\ref{eq:Lagrangian_Ito}) are
\emph{equivalent} in form to the standard equations of Brownian Dynamics
(BD), which are commonly used to model dynamics of colloidal particles
or polymer chains in flow \citet{BrownianDynamics_OrderN,LBM_vs_BD_Burkhard}.
In the absence of external forces, BD is typically presented as solving
the Ito equations of motion for the (correlated) positions of the
$N$ tracers (Brownian walkers) $\V Q=\left\{ \V q_{1},\dots,\V q_{N}\right\} $,
\begin{equation}
d\V Q=\left(2k_{B}T\,\M M\right)^{\frac{1}{2}}d\V{\mathcal{B}}+k_{B}T\left(\partial_{\V Q}\cdot\M M\right)dt+\sqrt{2\chi_{0}}\, d\M{\mathcal{B}}_{\V q},\label{eq:BD_M}
\end{equation}
where $\M M\left(\V Q\right)$ is the \emph{mobility} block matrix
for the collection of particles \citet{SELM}. This is equivalent
to (\ref{eq:Lagrangian_Ito}) with the identification of the mobility
tensor for a pair of particles $i$ and $j$,
\[
\M M_{ij}\left(\V q_{i},\V q_{j}\right)=\frac{1}{2k_{B}T}\sum_{k}\V{\phi}_{k}\left(\V q_{i}\right)\V{\phi}_{k}\left(\V q_{j}\right)=\frac{\M{\mathcal{R}}\left(\V q_{i},\V q_{j}\right)}{2k_{B}T}.
\]
If we write this explicitly for Stokes flow using (\ref{eq:v_cov_Stokes})
we get (see, for example, Eq. (3.25) in Ref. \citet{SIBM_Brownian}),
\begin{equation}
\M M_{ij}\left(\V q_{i},\V q_{j}\right)=\eta^{-1}\int\M{\sigma}\left(\V q_{i},\V r^{\prime}\right)\M G\left(\V r^{\prime},\V r^{\prime\prime}\right)\M{\sigma}^{T}\left(\V q_{j},\V r^{\prime\prime}\right)d\V r^{\prime}d\V r^{\prime\prime}.\label{eq:M_ij}
\end{equation}
When the particles are far apart, $\norm{\V q_{i}-\V q_{j}}\gg\sigma$,
the mobility is well-approximated by the Oseen tensor, $\M M_{ij}\left(\V q_{i},\V q_{j}\right)\approx\eta^{-1}\M G\left(\V q_{i},\V q_{j}\right)$.
At short distances the divergence of the Oseen tensor is mollified
by the filter, and (\ref{eq:M_ij}) gives a pairwise mobility very
similar to the Rotne-Prager-Yamakawa (RPY) mobility used in BD simulations
\citet{ISIBM}. In future work we will show that one can start from
the overdamped Lagrangian equations (\ref{eq:Lagrangian_Ito}), or
equivalently, (\ref{eq:BD_M}), and follow an argument similar to
the one of Dean \citet{SPDE_Diffusion_Dean}, which omits hydrodynamic
correlations, to construct an overdamped Eulerian equation for the
empirical concentration $c\left(\V r,t\right)=\sum_{i=1}^{N}\delta\left(\V q_{i}\left(t\right)-\V r\right)$.
Such a calculation leads to the same equation (\ref{eq:limiting_Ito})
as we obtained here by rather different means, and demonstrates that
hydrodynamic interactions do not modify Fick's law (\ref{eq:dc_dt_mean})
for the mean concentration. It has been suggested that non-local diffusion
terms appear when hydrodynamis is accounted for \citet{DDFT_Lowen,DDFT_Hydro_Lowen};
however, this prior work fails to notice that for the RPY mobility
these terms actually disappear as a direct consequence of the incompressibility
of the fluid flow.

In principle, traditional BD can be used to study the Lagrangian tracer
dynamics numerically. This has in fact been done by some authors in
turbulence to study multi-particle correlations of a \emph{few} passive
tracers \citet{Kraichnan_Lagrangian}. A key difference is that in
traditional BD the stochastic terms are generated by applying some
form of square root of the mobility $\M M\left(\V Q\right)$, which
can be expensive for many tracers unless specialized fast multipole
techniques are employed \citet{BrownianDynamics_OrderN,BrownianDynamics_OrderNlogN}.
By contrast, in the equivalent formulation (\ref{eq:Lagrangian_Ito})
the stochastic forcing is generated by evaluating a random velocity
field at the positions of the tracers. This formulation leads to a
simple Lagrangian algorithm that is \emph{linear} in the number of
tracers $N$, as we discuss in Section \ref{sub:Algorithm}.

\subsection{Relation to mode-mode coupling and renormalization theories}

The fact that thermal velocity fluctuations enhance diffusion is well
known and there are several mode-mode coupling calculations that eventually
lead to a similar result to our Stokes-Einstein formula (\ref{eq:chi_r_Stokes}).
A key difference between our approach and other derivations is the
fact that our calculation replaced typical \emph{uncontrolled} approximations
by a precise set of initial assumptions, and leads to a \emph{rigorous}
closed-form fluctuating advection-diffusion equation (\ref{eq:Lagrangian_Ito})
for the concentration. In traditional perturbative renormalization
approaches \citet{DiffusionRenormalization_I,DiffusionRenormalization_II,DiffusionRenormalization_III,ExtraDiffusion_Vailati,DiffusionRenormalization},
one starts from equations that already have diffusion (dissipation)
in them, and then considers what perturbation the fluctuations make.
In this sense, Fick's linear law is the zeroth order approximation,
and the first order perturbation is \emph{linearized} fluctuating
hydrodynamics. At the next order the fluctuations are found to give
rise to Fick's law with a renormalized diffusion obeying a Stokes-Einstein
relation \citet{ExtraDiffusion_Vailati}. This leads to a circular
argument in which the physical phenomenon included in the lowest order
approximation is the result of higher-order approximations, and an
infinite sequence of renormalization steps is required to make the
model self-consistent. Our work shows that the problem is quite simply
and straightforwardly solved by starting with a \emph{non-linear model}
that then self-consistently gives rise to ``renormalized'' diffusion,
instead of starting with a model that has diffusion put in as input
and then linearizing.

While the physical difference between two and three dimensional fluctuations
has been long appreciated in the literature, we believe our approach
is not only simpler, but also more effective and more illuminating
than mode-mode coupling analysis. There has been some confusion in
the literature about the applicability of hydrodynamics to two dimensional
systems, and statements to the effect that Stokes-Einstein does not
apply in two dimensions have been made \citet{SE_2D}. We showed that
for finite systems nonlinear fluctuating hydrodynamics does lead to
Stokes-Einstein relation for the diffusion coefficient in Fick's law
(\ref{eq:dc_dt_mean}) for the ensemble \emph{mean}. One of the reasons
we are able to easily obtain results is our use of the separation
of time scales. We note, however, that the assumption of infinite
Schmidt number, crucial to our approach, has to fail in \emph{very}
large two dimensional systems. Namely, large-scale (slow) velocity
modes make a crucial contribution to diffusion, which itself occurs
at faster time scales as the system grows due to the increasing diffusion
coefficient. At finite Schmidt numbers the situation is much more
complex and even mode-mode coupling theories run into problems \citet{BilinearModeCoupling_SE}.
The asymptotic behavior of the system of equations (\ref{eq:v_eq},\ref{eq:c_eq_original})
in two dimensions in the infinite system size (thermodynamic) limit
remains an interesting open question \citet{StokesEinstein}.

\subsection{\label{sub:Algorithm}Multiscale Numerical Algorithms}

Details of our multiscale numerical algorithms for solving the limiting
Eulerian (\ref{eq:limiting_Strato}) and Lagrangian (\ref{eq:Lagrangian_Strato})
equations are given in Appendix \ref{sec:TemporalScheme}; here we
briefly summarize the key features.

We have developed finite-volume numerical methods to simulate the
limiting dynamics (\ref{eq:limiting_Strato}), detailed in Section
\ref{sub:EulerianAlgorithm}. The spatial discretization of the advective
term $\V w\odot\grad c$ is identical to the one described in Ref.
\citet{LLNS_Staggered}, and is constructed to ensure that advection
is discretely non-dissipative. A pseudo-spectral steady-Stokes solver
is used to generate a random advection velocity $\V w$. The temporal
integrator uses the implicit midpoint rule for the term $\chi_{0}\grad^{2}c$
\citet{LLNS_Staggered}, and the Euler-Heun method %
\footnote{The Euler-Heun method is a predictor-corrector algorithm that can
be can be thought of as the Stratonovich equivalent of the Euler-Maruyama
method for Ito stochastic differential equations.%
} for the term $\V w\odot\grad c$. This approach is chosen because
it ensures fluctuation-dissipation balance between the enhanced diffusion
and the random advection. Note that in this numerical method the small-scale
cutoff $\sigma$ is related to the grid spacing employed in the finite-volume
grid. The results of our numerical algorithm are compared to the results
for the resolved dynamics (\ref{eq:v_eq},\ref{eq:c_eq_original})
in Fig. \ref{fig:DiffusiveInterface}. Visually the two figure panels
are indistinguishable, and more detailed analysis has not found any
statistically significant differences. It is, however, important to
point out that the multiscale method employing the limiting dynamics
can reach the same time scales in \emph{much} less computational effort
than the direct numerical simulation because it avoids the need to
resolve the fast velocity fluctuations. Because it plays little role
far from equilibrium and does not affect the giant fluctuations that
are the focus of our study, we do not include the multiplicative noise
term $\grad\cdot\left(\sqrt{2\chi_{0}c}\,\M{\mathcal{W}}_{c}\right)$
from the concentration equation in all of the Eulerian numerical simulations
reported here.

We cannot study the case of no bare diffusion, $\chi_{0}=0$, using
an \emph{Eulerian} grid-based algorithm. Because advection creates
finer and finer scales in the solution, truncation on a regular grid
leads to a Gibbs phenomenon and ultimately numerical instability.
We are, in fact, not aware of any Eulerian numerical method that could
be used to reliably study the limiting case $\chi_{0}=0$ without
introducing artificial dissipation. We have therefore developed a
\emph{Lagrangian} tracer algorithm to solve (\ref{eq:Lagrangian_Strato}),
detailed in Section \ref{sub:LagrangianAlgorithm}. In the Lagrangian
algorithm, a realization of the advection velocity field $\V w\left(\V r,t\right)$
is sampled using a spectral steady Stokes solver, a convolution with
a Gaussian function of standard deviation $\sigma$ is used to filter
the small scales, and a non-uniform FFT algorithm \citet{NUFFT} is
used to evaluate $\V w(\V q,t)$. This approach leads to a scheme
in which the only truncation (discretization) error comes from the
Euler-Heun temporal integrator.

The Lagrangian approach is particularly useful when the tracers model
actual physical particles that can be tracked individually, for example,
fluorescently-labeled molecules in a FRAP experiment. The Lagrangian
tracing algorithm can also be used to solve (\ref{eq:limiting_Strato},\ref{eq:limiting_Ito})
in the abscence of bare diffusion by employing the identity $c\left(\V q(t),t\right)=c\left(\V q(0),0\right)$.
Because the Lagrangian trajectories are time-reversible, one can obtain
the concentration at a given position $\V r$ by starting a tracer
from $\V q(0)=\V r$, following its trajectory for a time $t$, and
then evaluating the initial condition at the new position of the tracer,
$c\left(\V q(0),t\right)=c\left(\V q(t),0\right)$. The Lagrangian
approach leads to a spatial discretization of (\ref{eq:limiting_Strato})
that is free of artificial dispersion or dissipation, with the main
source of numerical error coming from the fact that a finite number
of tracers is employed.

\section{\label{sec:Irreversibility}Is Diffusion in Liquids Irreversible?}

The well-known fact that the measured diffusion coefficients in molecular
liquids and macromolecular solutions closely match the Stokes-Einstein
prediction hints that in realistic fluids diffusive transport is dominated
by advection by the velocity fluctuations, $\chi\gg\chi_{0}$. This
suggests that it is relevant to consider the case of no bare diffusion.
If $\chi_{0}=0$, the evolution of the mean is dissipative since $\chi_{\text{eff}}=\chi>0$.
However, each realization follows a strictly reversible dynamics.
It is not difficult to appreciate that an instance of a random process
can behave very differently from the ensemble mean. For example, the
average over many randomly dephasing oscillators will produce a decaying
amplitude, even though each instance is non-decaying. It is therefore
important to understand the difference in the behavior of the ensemble
mean of the diffusive mixing process, described by (\ref{eq:dc_dt_mean}),
and the behavior of an individual realization, described by (\ref{eq:limiting_Strato}). 

\begin{figure*}[tbph]
\begin{centering}
\includegraphics[width=0.31\textwidth]{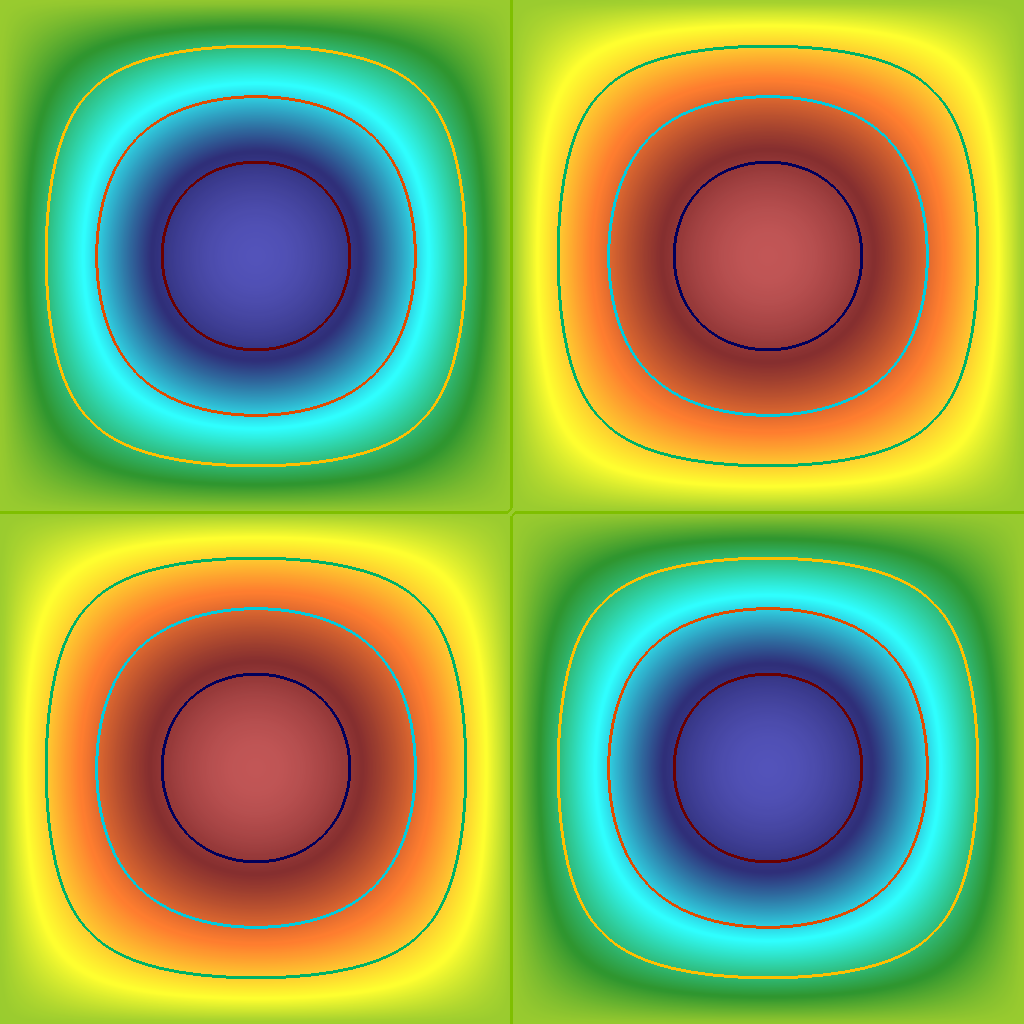}\hspace{0.2cm}\includegraphics[width=0.31\textwidth]{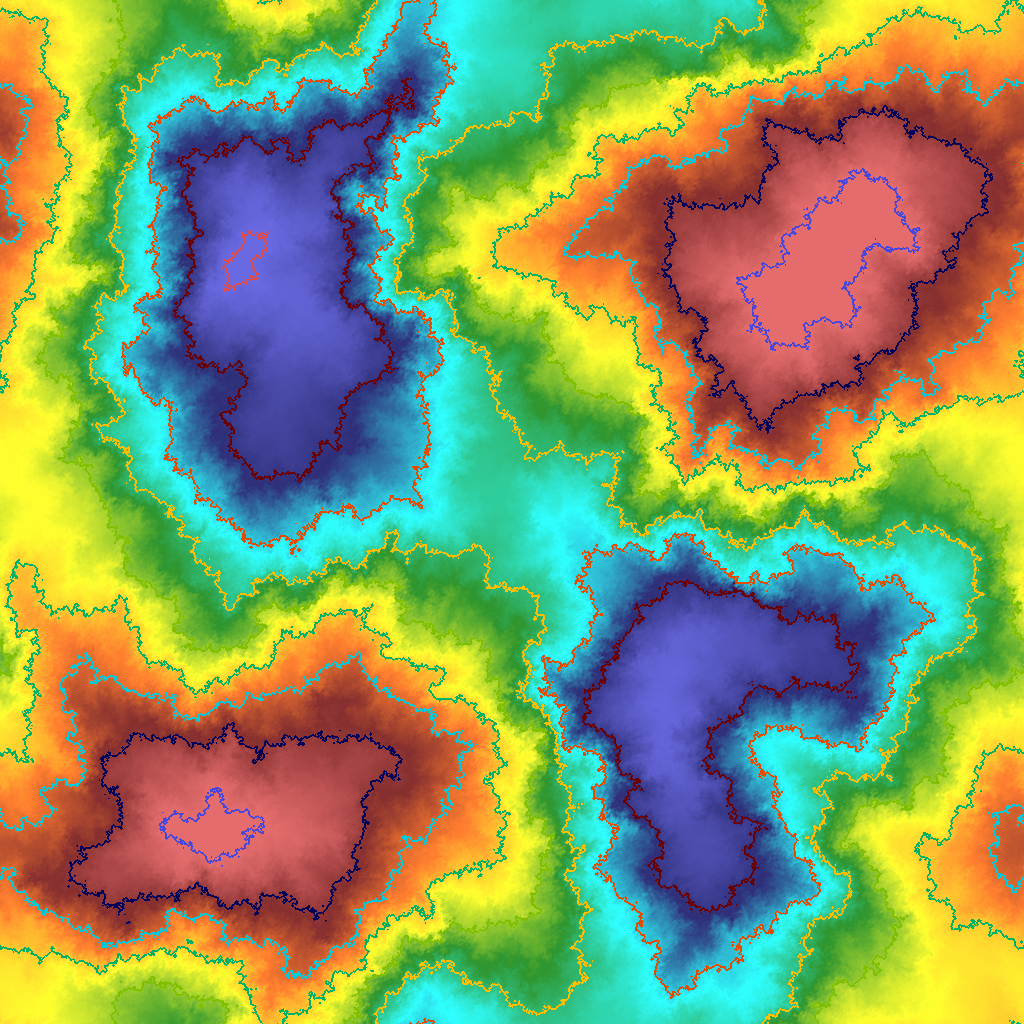}\hspace{0.2cm}\includegraphics[width=0.31\textwidth]{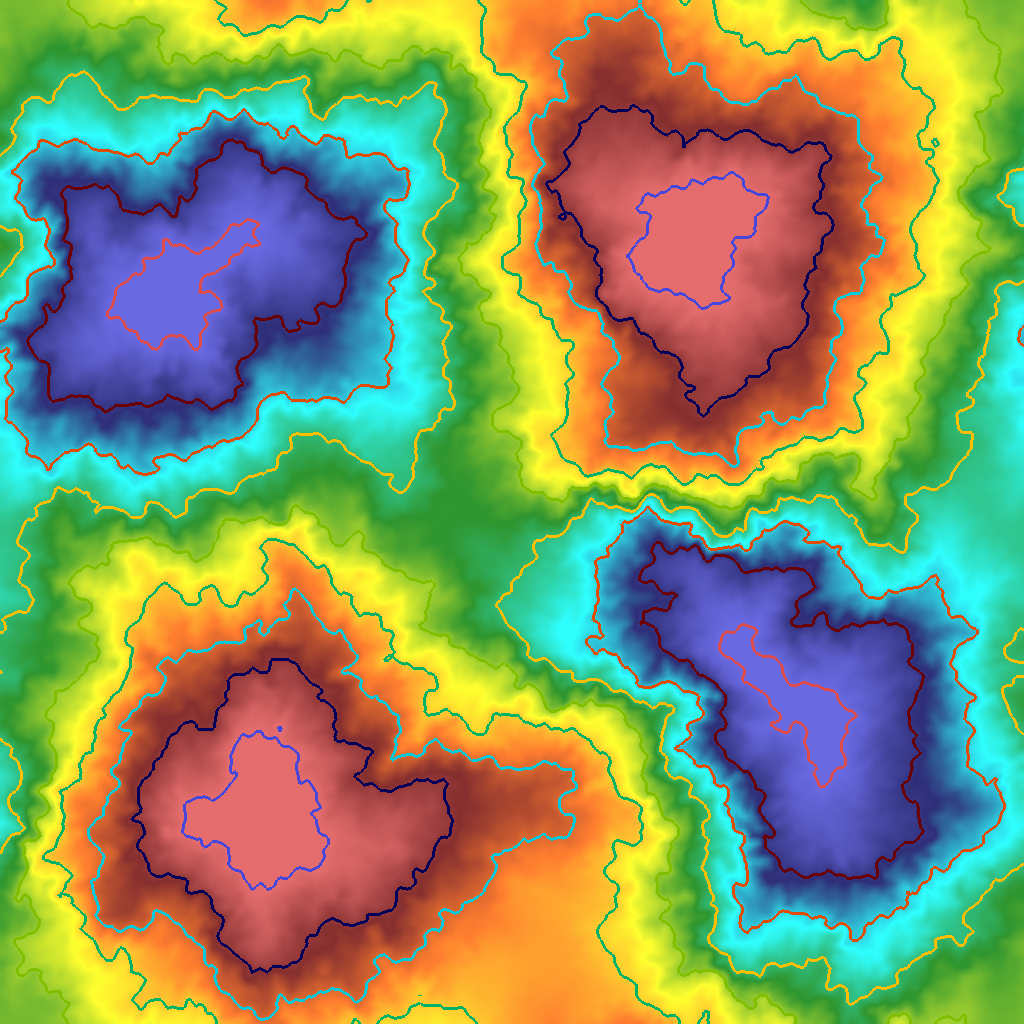}
\par\end{centering}

\caption{\label{fig:SmoothDecay}The concentration obtained after a substantial
decay of a single-mode initial condition $c(\V r,0)=\sin\left(2\pi x/L\right)\sin\left(2\pi y/L\right)$.
Contour lines are also shown, and the same final time $t\sim\tau$
and color legend is used in all three panels. (\emph{Left panel})
The ensemble averaged mean, which follows (\ref{eq:dc_dt_mean}).
(\emph{Middle}) Solution of the stochastic advection-diffusion equation
(\ref{eq:limiting_Strato}) on a grid of size $1024\times1024$ cells
for the case $\chi_{\text{eff}}/\chi_{0}\approx50$. (\emph{Right
panel}) An instance of the solution of (\ref{eq:limiting_Strato})
for a grid of size $256\times256$ cells, for $\chi_{\text{eff}}/\chi_{0}\approx5$,
with the same $\chi_{\text{eff}}=\chi_{0}+\chi$ as the other two
panels.}
\end{figure*}

To this end, let us consider the temporal decay of a smooth single-mode
initial perturbation $c(\V r,0)=\sin\left(2\pi x/L\right)\sin\left(2\pi y/L\right)$
in two dimensions, using our numerical method for simulating the overdamped
dynamics (\ref{eq:limiting_Strato},\ref{eq:limiting_Ito}). In the
left panel of Fig. \ref{fig:SmoothDecay}, we show the ensemble mean
of the concentration at a later time, as obtained by solving the deterministic
equation (\ref{eq:dc_dt_mean}) with an effective diffusion coefficient
$\chi_{\text{eff}}=\chi_{0}+\chi$, where the value of $\chi$ was
obtained via the discrete equivalent of (\ref{eq:chi_r_Stokes}).
In the middle panel of the figure, we show an instance of the concentration
at the same time obtained by solving (\ref{eq:limiting_Strato},\ref{eq:limiting_Ito})
using the smallest value of $\chi_{0}$ that stabilized the numerical
scheme. The same giant fluctuations seen in Fig. \ref{fig:DiffusiveInterface}
are revealed, with the contour lines of the concentration becoming
rough even at the scale of the grid spacing. We note in passing that
we have performed hard-disk molecular dynamics simulation of this
mixing process and have observed the same qualitative behavior seen
in the middle panel of Fig. \ref{fig:SmoothDecay}.

\subsection{Power Transfer}

The conserved quantity%
\footnote{Advection preserves not just the second but all moments of the concentration.%
} $\int\left(c^{2}/2\right)d\V r$ injected via the initial perturbation
away from equilibrium is effectively dissipated through a mechanism
similar to the energy cascade observed in turbulent flows. Advection
transfers power from the large length scales to the small length scales,
\emph{effectively} dissipating the power injected into the large scales
via the initial condition. To make this more quantitative, let us
set $\chi_{0}=0$ and write (\ref{eq:limiting_Strato}) in the Fourier
domain,
\[
\frac{d\hat{c}_{\V k}}{dt}=\hat{\V w}_{\V k}\circledast\left(i\V k\hat{c}_{\V k}\right),
\]
where $\circledast$ is a combination of Stratonovich dot product
and convolution. This equation strictly conserves the total power
$\sum_{\V k}\abs{\hat{c}_{\V k}\left(t\right)}^{2}/2$ since the advective
term simply redistributes the power between the modes. Let us denote
the ensemble average power in mode $\V k$ with $p_{\V k}\left(t\right)=\av{\abs{\hat{c}_{\V k}\left(t\right)}^{2}}/2$.
Using straightforward stochastic calculus we can obtain a simple system
of ODEs for the transfer of power between the modes,
\begin{equation}
\frac{dp_{\V k}}{dt}=-\sum_{\V k^{\prime}\neq\V k}\left(\V k\cdot\hat{\M{\mathcal{\chi}}}_{\V k-\V k^{\prime}}\cdot\V k\right)p_{\V k}+\sum_{\V k^{\prime}\neq\V k}\left(\V k^{\prime}\cdot\hat{\M{\mathcal{\chi}}}_{\V k-\V k^{\prime}}\cdot\V k^{\prime}\right)p_{\V k^{\prime}}.\label{eq:dp_k_dt}
\end{equation}
The first term on the right hand side of this equation expresses the
power lost from mode $\V k$ to other modes, while the second term
gives the power transferred from other modes to mode $\V k$. The
total power $\sum_{\V k}p_{\V k}$ is conserved because the flow in
(\ref{eq:Lagrangian_Strato}) is unique, ensuring that there is no
anomalous dissipation \citet{AnomalousDissipation}.

Consider starting from an initial configuration in which the only
mode with nonzero power is wavenumber $\V k_{0}$, $\hat{c}_{\V k}\left(0\right)=\delta_{\V k,\V k_{0}}$.
The average rate at which power will be transferred from mode $\V k_{0}$
to mode $\V k\neq\V k_{0}$ via the advective term $-\V w\odot\grad c$
is proportional to the spectrum of $\V w$ at wavenumber $\V k-\V k_{0}$
and is given by $\V k_{0}\cdot\hat{\M{\mathcal{\chi}}}_{\V k-\V k_{0}}\cdot\V k_{0}$.
The total relative rate at which power is lost (``dissipated'')
from mode $\V k_{0}$ is given by $\V k_{0}\cdot\M{\chi}\cdot\V k_{0}$,
where $\M{\chi}=\sum_{\V k}\hat{\M{\mathcal{\chi}}}_{\V k}$ for a
finite system case, or $\M{\chi}=\left(2\pi\right)^{-d}\int\hat{\M{\mathcal{\chi}}}_{\V k}d\V k$
in the infinite system limit. This is exactly the same rate of dissipation
as one would get for ordinary diffusion with diffusion tensor $\M{\chi}$.
In simple diffusion, the power of mode $\V k_{0}$ would also decay
exponentially as $\exp\left(-t/\tau\right)$, where $\tau=\left(2\chi_{\text{eff}}k_{0}^{2}\right)^{-1}$
is a decay time. However, all other modes would remain unexcited,
as in the left panel of Fig. \ref{fig:SmoothDecay}.

\begin{figure}[tbph]
\begin{centering}
\includegraphics[width=0.75\columnwidth]{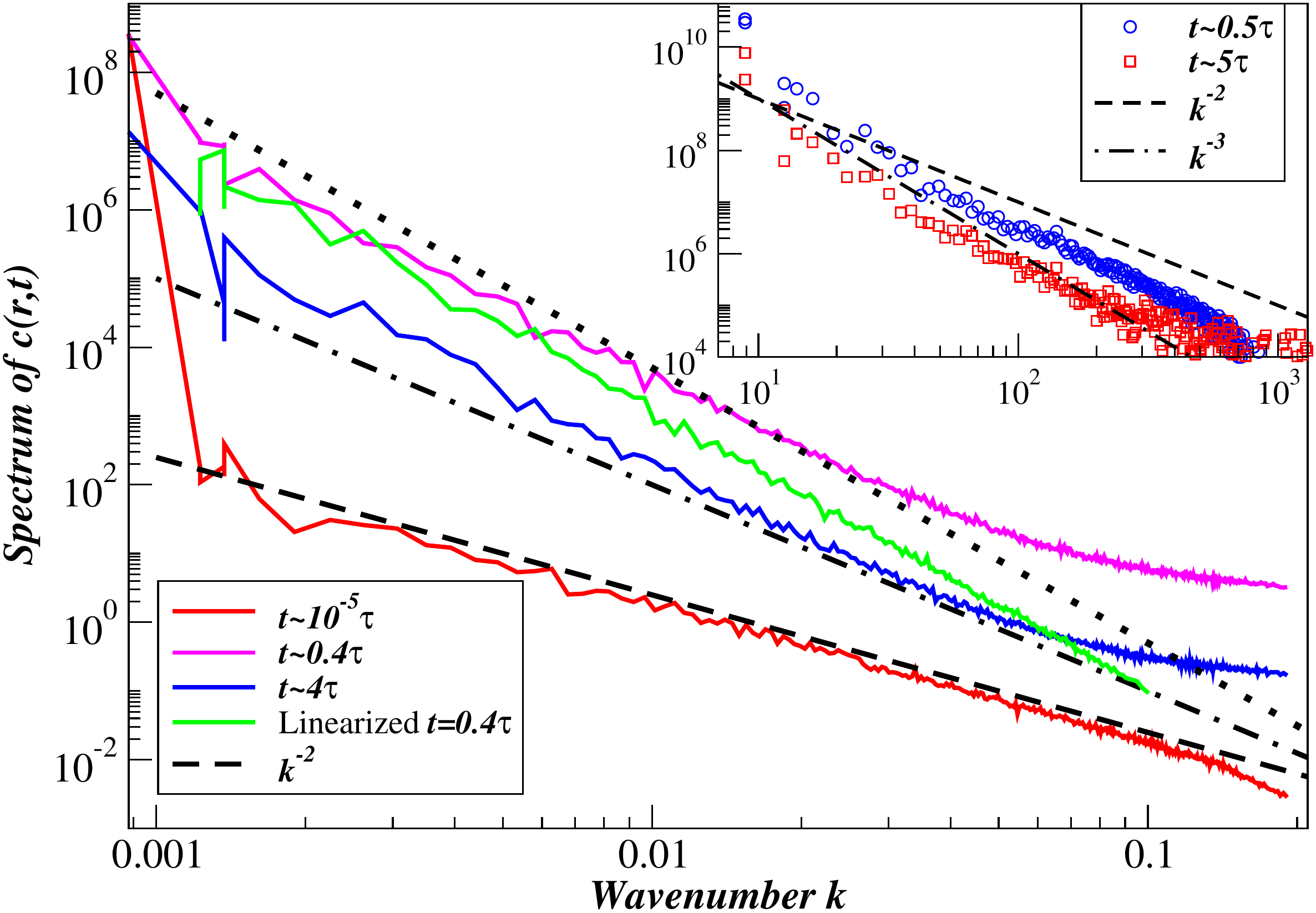}
\par\end{centering}

\caption{\textbf{\label{fig:S_k_c}}Power spectrum of an \emph{individual}
realization of the concentration $c(\V r,t)$ corresponding to the
simulation illustrated in the middle panel in Fig. \ref{fig:SmoothDecay}.
The power of individual modes $\V k$ with nearby $k$ is averaged
and the result is shown with a colored solid line, while dashed/dotted
lines show power laws $k^{-2}$, $k^{-3}$ and $k^{-4}$ for comparison.
At early times $t\ll\tau=\left(2\chi_{\text{eff}}k_{0}^{2}\right)^{-1}$
(red line) power is being transferred from mode $k_{0}\approx2\pi/L\approx10^{-3}$,
initially excited to have spectral power $p_{\V k_{0}}\approx7\cdot10^{8}$
(off scale), to the rest of the modes, leading to a spectrum $\sim k^{-2}$.
At late times $t\gtrsim\tau$ (magenta and blue lines), a steadily-decaying
shape of the spectrum is reached where power transferred from the
larger scales is dissipated at the small scales via bare diffusion.
Linearized fluctuating hydrodynamics predicts a spectrum $\sim k^{-4}$
(green line). (\emph{Inset}) Difference in the spectrum between random
advection and simple diffusion, as obtained using a Lagrangian simulation
of the diffusive decay. The parameters used are different from the
main panel and are summarized in Section \ref{sec:SpatialCG}.}
\end{figure}

The above calculation shows that after a short time $t\ll\tau$, mode
$\V k\neq\V k_{0}$ will have, on average, power proportional to $\left(\V k_{0}\cdot\hat{\M{\mathcal{\chi}}}_{\V k-\V k_{0}}\cdot\V k_{0}\right)t$.
For large-scale initial pertubations, $\V k_{0}\approx\V 0$, the
spectrum of $c$ at short times will therefore be proportional to
one of the diagonal elements of $\hat{\M{\mathcal{\chi}}}_{\V k}$,
which can be read from (\ref{eq:chi_k_filtered}) to be $\sim k^{-2}\sin^{2}\theta$
for intermediate wavenumbers, where $\theta$ is the angle between
$\V k$ and $\V k_{0}$. The spectrum at early times is therefore
$\abs{\hat{c}_{\V k}\left(t\right)}^{2}\sim k^{-2}t$, as confirmed
by our numerical simulations and illustrated in Fig. \ref{fig:S_k_c}.

If there were only random advection, with no bare diffusion, the transfer
of energy from the coarse to the fine scales would continue indefinitely.
It is not hard to see that even a very small finite bare diffusion
can affect the results at small scales dramatically, making the limit
$\chi_{0}\rightarrow0^{+}$ non-trivial \citet{Kraichnan_NoDiffusion}.
Namely, the diffusive term $\chi_{0}\grad^{2}c$ becomes stronger
and stronger at smaller scales ($\chi_{0}k^{2}$ in Fourier space),
and will eventually become important and dissipate the small scale
features created by the random advection. In particular, at late times
of the diffusive decay, $t\sim\tau$, shown in Fig. \ref{fig:SmoothDecay},
one expects that a steadily decaying state will be reached in which
the shape of the spectrum of $c$ does not change as it decays exponentially
in time as $\exp\left(-t/\tau\right)$. This is indeed what we observe,
and the shape of the steadily decaying spectrum is shown in Fig. \ref{fig:S_k_c}.
Numerically we observe that the majority of the bare dissipation occurs
at the largest wavenumbers, dissipating the power injected into small
scales from the large and intermediate scales. This is seen in Fig.
\ref{fig:S_k_c} as a deviation from the power-law behavior for the
largest wavenumbers. It is important to note that the shape of the
spectrum at the large wavenumbers is strongly affected by discretization
artifacts for the finite volume scheme employed here.

We also study the spectrum of concentration fluctuations with the
Lagrangian tracer algorithm described in Section \ref{sub:Algorithm},
which allows us to eliminate bare diffusion and numerical grid artifacts.
Note however that the Lagrangian approach to obtaining power spectra
also fails at sufficiently large wavenumbers at sufficiently large
times, as we explain shortly. In the Lagrangian algorithm, we make
use of the time-reversibility of the flow, $c\left(\V q(0),t\right)=c\left(\V q(t),0\right)$.
We place $N_{t}=2048^{2}$ on a regular grid at the initial time and
then follow their trajectories over a time interval $t$ using the
algorithm described in Section \ref{sub:LagrangianAlgorithm}. We
then evaluate the initial condition $c\left(\V r,0\right)$ at the
final position of each tracer in order to obtain the concentration
$c\left(\V r,t\right)$ on a regular grid of points, and use the FFT
algorithm to obtain the spectrum $\hat{c}_{\V k}\left(t\right)$.
This would be a very accurate numerical algorithm if the concentration
were smooth on the scale of the grid of tracers. In reality, as the
power-law tail in the spectrum gets filled by the advection the concentration
becomes less and less smooth and the spectrum at the larger wavenumbers
becomes dominated by truncation errors (using a discrete sum instead
of a Fourier integral) and statistical errors (using a finite number
of sampling points to obtain the spectrum).

In order to eliminate these artifacts and further emphasize the difference
between simple diffusion and advection by a random field, we consider
repeating the Lagrangian calculation with tracers that perform \emph{independent}
Brownian motions with diffusion coefficient $\chi_{\text{eff}}$.
For simple diffusion, the numerical spectrum is not zero for $\V k\neq\V k_{0}$
as it should be; rather, due to the finite number of tracers we get
$p_{\V k}\left(t\right)\sim N^{-1}\left[1-\exp\left(-t/\tau\right)\right]$.
We subtract this background noise from the numerical spectrum obtained
using the Lagrangian tracing algorithm. We find that the difference
in the spectrum for random advection and simple diffusion follows
a power-law behavior, as illustrated in the inset of Fig. \ref{fig:S_k_c}.
The power law is in agreement with that seen in the Eulerian simulations,
and persists over the whole range of accessible wavenumbers.

\subsection{\label{sub:Linearized}Linearized Fluctuating Hydrodynamics}

In the literature, linearized fluctuating hydrodynamics is frequently
used to obtain the steady-state spectrum of fluctuations \citet{FluctHydroNonEq_Book}.
In the limit of large Schmidt numbers, the standard heuristic approach
leads to the additive-noise equation,
\begin{equation}
\partial_{t}\tilde{c}=-\V w\cdot\grad\av c+\grad\cdot\left[\chi_{\text{eff}}\grad\tilde{c}\right],\label{eq:limiting_linearized}
\end{equation}
where $\av c$ is the ensemble mean, which follows (\ref{eq:dc_dt_mean}).
Note that in order to obtain the correct spectrum for the \emph{equilibrium}
concentration fluctuations, one ought to include an additional random
forcing term $\grad\cdot\left(\sqrt{2\chi_{\text{eff}}\av c}\,\M{\mathcal{W}}_{c}\right)$
in (\ref{eq:limiting_linearized}); our focus here is on the \emph{nonequilibrium}
fluctuations and we will not include such a term to more accurately
measure the power-law spectrum. Equation (\ref{eq:limiting_linearized})
can easily be solved analytically in the Fourier domain when $\grad\av c=\V h$
is a weak externally applied constant gradient (c.f., for example,
Eq. (9) in Ref. \citet{DiffusionRenormalization}), to obtain a spectrum
$\left(\V h\cdot\hat{\M{\mathcal{\chi}}}_{\V k}\cdot\V h\right)/\left(\chi_{\text{eff}}k^{2}\right)\sim k^{-4}$
for intermediate wavenumbers. 

Linearized fluctuating hydrodynamics can be used to justify the omission
of the term $\grad\cdot\left(\sqrt{2\chi_{0}c}\,\M{\mathcal{W}}_{c}\right)$
from (\ref{eq:c_eq_original}) in our numerical calculations. This
term leads to ``equilibrium'' concentration fluctuations that are
negligible compared to the ``non-equilibrium'' concentration fluctuations
due to advection by the thermal velocity fluctuations in the presence
of large concentration gradients. A rough estimate of the magnitude
of the equilibrium versus the nonequilibrium fluctuations can be obtained
by considering the case when the tracer particles are labeled molecules
of a simple three-dimensional liquid with molecular mass $m$, molecular
diameter $\sigma$ and molecular number density $n=\rho/m\sim\sigma^{-3}$.
The spectrum of the equilibrium fluctuations is $\av{\abs{\hat{c}_{\V k}}^{2}}_{\text{eq}}\approx\av c\sim\phi\sigma^{-3}$,
where $\phi$ is the volume (packing) fraction of the labeled particles.
In the presence of a constant applied concentration gradient of magnitude
$h$, an additional nonequilibrium contribution $\av{\abs{\hat{c}_{\V k}}^{2}}_{\text{neq}}\approx h^{2}n^{2}k_{B}T/\left(\eta\chi_{\text{eff}}k^{4}\right)$
is obtained (see Appendix A in Ref. \citet{LowMachExplicit}, but
note that in that work $c$ denotes mass fraction, rather than number
concentration as it does in this paper). To estimate the ratio of
these two contributions we can use the Stokes-Einstein relation $\chi_{\text{eff}}\sim k_{B}T/\left(\eta\sigma\right)$
and $h\sim L^{-1}$, where $L\gg\sigma$ is the scale of the applied
gradient, to obtain
\[
\av{\abs{\hat{c}_{\V k}}^{2}}_{\text{eq}}/\av{\abs{\hat{c}_{\V k}}^{2}}_{\text{neq}}\sim\phi\, L^{2}\sigma^{2}k^{4}.
\]
At length scales $k\sim L^{-1}$, the above ratio is $\sim\left(\sigma/L\right)^{2}\ll1$,
and in practical situations the nonequilibrium fluctuations are ``giant''
compared to the equilibrium ones.

For finite gradients and more realistic boundary conditions, we can
solve (\ref{eq:limiting_linearized}) numerically with the same algorithm
used to solve the full nonlinear equation (\ref{eq:limiting_Strato}).
Namely, we set $\chi_{0}=\chi_{\text{eff}}$ and reduce the magnitude
of the fluctuations by a factor $\epsilon\ll1$ by setting the ``temperature''
to $\epsilon k_{B}T$, and then simply rescale the spectrum of the
fluctuations by a factor $\epsilon^{-1}$ to obtain the spectrum of
$\tilde{c}$. This approach was used by some of us to simulate giant
concentration fluctuations in microgravity \citet{FractalDiffusion_Microgravity}
in Ref. \citet{LLNS_Staggered}. The result of this numerically-linearized
calculation for the single-mode initial condition is shown in Fig.
\ref{fig:S_k_c} and compared to the nonlinear calculations. The spectrum
is indeed seen to follow a power-law $k^{-4}$ for the linearized
equations, in agreement with theory. Note however that the spectrum
obtained from (\ref{eq:limiting_linearized}) is not in a very good
match with the spectrum obtained by solving (\ref{eq:limiting_Strato}),
which appears closer to $k^{-3}$ in the two-dimensional setting we
study here. Tools developed in the turbulence literature \citet{Kraichnan_SelfSimilar,Kraichnan_SelfSimilarPRL}
could potentially be used to study the spectrum of uniformly decaying
steady states without resorting to linearization. Alternatively, the
system of differential equations (\ref{eq:dp_k_dt}) can be solved
numerically to study the average dynamics of the transfer of power
between the modes.

By integrating the $\sim k^{-4}$ spectrum of concentration fluctuations
predicted by (\ref{eq:limiting_linearized}) it can easily be seen
that in two dimensions the fluctuations of the concentration around
the Fickian mean are on the order of the applied concentration gradient.
Therefore, they \emph{cannot} be considered ``microscopic'' or ``small'',
and linearized fluctuating hydrodynamics does \emph{not} apply in
two dimensions. For example, the solution of (\ref{eq:limiting_linearized})
does not necessarily stay positive due to the large fluctuations,
as we have observed numerically for parameters representative of moderately-dense
hard-disk systems. This is an inherent pitfall of linearizing the
nonlinear advective term when fluctuations become truly ``giant''
(as they do in two dimensions). By contrast, the nonlinear equations
preserve the bounds on concentration even when the fluctuations become
strong, since advecting by a spatially smooth (even if white in time)
velocity obeys a monotonicity principle.

\section{\label{sec:SpatialCG}Spatial Coarse-Graining}

If there were only random advection, with no bare diffusion, the transfer
of energy from the coarse to the fine scales would continue indefinitely,
since the dynamics is reversible and there is nothing to dissipate
the power. However, any features in $c$ at length scales below molecular
scales have no clear physical meaning. In fact, continuum models are
inapplicable at those scales. It is expected that not resolving (coarse-graining)
the microscopic scales will lead to true dissipation and irreversibility
in the coarse-grained dynamics. Such coarse-graining can take form
of ensemble averaging, or elimination of slow degrees of freedom.
In either case, the loss of knowledge about the small scales will
lead to positive entropy production.

Can one replace the molecular scale details, or even all details of
the dynamics at scales below some mesoscopic observation scale $\delta$,
by some simple approximation, for example, a bare diffusion term with
suitably chosen $\chi_{0}$? We propose here a way to carry out such
\emph{spatial} coarse-graining of the overdamped dynamics (\ref{eq:limiting_Strato},\ref{eq:limiting_Ito})
by splitting the velocity $\V w$ into a large-scale component $\V w_{\delta}$
and a small-scale component $\tilde{\V w}$,
\[
\V w=\M{\delta}\star\V w+\tilde{\V w}=\V w_{\delta}+\tilde{\V w},
\]
where $\M{\delta}$ is a filter that smooths scales below some mesoscopic
length $\delta>\sigma$. More precisely, the equality $\V w=\V w_{\delta}+\tilde{\V w}$
is in law and corresponds to splitting the covariance matrix $\M{\mathcal{R}}=\M{\delta}\star\M{\mathcal{R}}\star\M{\delta}^{T}+\widetilde{\M{\mathcal{R}}}$
into a small-scale and large-scale component, and generating the two
parts of $\V w$ independently. Because of the technical difficulty
in dealing with the multiplicative noise term $\grad\cdot\left(\sqrt{2\chi_{0}c}\,\M{\mathcal{W}}_{c}\right)$
in (\ref{eq:limiting_Ito}), we do not include this term in this analysis.

In Eq. (\ref{eq:dc_dt_mean}), we performed an ensemble average over
all realizations of $\V w$. We can also, however, only average over
realizations of the unresolved $\tilde{\V w}$, that is, we can define
$\bar{c}_{\delta}=\av c_{\tilde{\V w}}$ as the conditional ensemble
average keeping $\V w_{\delta}$ fixed. We can directly take such
a conditional average of the Ito equation (\ref{eq:limiting_Ito}),
to obtain, \emph{without} any approximations, a closed equation for
$\bar{c}_{\delta}$ of exactly the same form as (\ref{eq:limiting_Ito}),
\begin{equation}
\partial_{t}\bar{c}_{\delta}=-\V w_{\delta}\cdot\grad\bar{c}_{\delta}+\chi_{0}\grad^{2}\bar{c}_{\delta}+\grad\cdot\left[\M{\chi}\left(\V r\right)\grad\bar{c}_{\delta}\right],\label{eq:filtered_c_Ito}
\end{equation}
with exactly the same initial condition, and, importantly, with an
identical effective diffusion coefficient $\chi_{\text{eff}}=\chi_{0}+\M{\chi}$.
However, if we write (\ref{eq:filtered_c_Ito}) in the Stratonovich
form used by our numerical methods, we see that the bare diffusion
coefficient needs to be \emph{renormalized} to take into account the
coarse-grained scales,
\begin{equation}
\partial_{t}\bar{c}_{\delta}=-\V w_{\delta}\odot\grad\bar{c}_{\delta}+\grad\cdot\left[\left(\chi_{0}+\D{\M{\chi}_{\delta}}\right)\grad\bar{c}_{\delta}\right],\label{eq:filtered_c_Strato}
\end{equation}
where the diffusion renormalization $\D{\M{\chi}_{\delta}}$ is
\begin{align}
\D{\M{\chi}_{\delta}}\left(\V r\right) & =\frac{1}{2}\widetilde{\M{\mathcal{R}}}\left(\V r,\V r\right)=\frac{1}{2}\M{\mathcal{R}}\left(\V r,\V r\right)-\frac{1}{2}\int\M{\delta}\left(\V r,\V r^{\prime}\right)\M{\mathcal{R}}\left(\V r^{\prime},\V r^{\prime\prime}\right)\M{\delta}^{T}\left(\V r,\V r^{\prime\prime}\right)d\V r^{\prime}d\V r^{\prime\prime}\label{eq:chi_renormalization}
\end{align}
Note that the renormalized \emph{bare} diffusion coefficient $\chi_{0}\left(\delta\right)=\chi_{0}+\D{\M{\chi}_{\delta}}$
in (\ref{eq:filtered_c_Strato}) is nonzero even if $\chi_{0}=0$.
This true dissipation is a remnant of the unresolved (eliminated)
small scales. This renormalized bare diffusion coefficient is not,
however, a material constant, but rather, depends on the mesoscopic
lengthscale $\delta$.

\subsection{Coarse-Grained Stochastic Advection-Diffusion Model}

In reality, we are not interested in the behavior of the conditional
average $\bar{c}_{\delta}$ because this is not a measurable quantity.
Rather, we are interested in the behavior of individual realizations
of the spatially coarse-grained concentration $c{}_{\delta}=\M{\delta}\star c$,
which can be measured by observing $c$ at scales larger than some
experimental resolution $\delta$. A physically reasonable coarse-grained
model can be obtained by \emph{assuming} that $c{}_{\delta}$ follows
the same equation as the conditional mean $\bar{c}_{\delta}$, 
\begin{equation}
\partial_{t}c{}_{\delta}\approx-\V w_{\delta}\odot\grad c{}_{\delta}+\grad\cdot\left[\left(\chi_{0}+\D{\M{\chi}_{\delta}}\right)\grad c{}_{\delta}\right],\label{eq:coarse_grained_Strato}
\end{equation}
so long as the initial condition is smooth %
\footnote{One can also use initial condition $c{}_{\delta}(0)=\M{\delta}\star c(0)$
since the enhanced bare diffusion quickly damps fine-scale features
in the initial condition.%
} at the scale $\delta$. It is important to note that, in order to
obtain correct equilibrium fluctuations, there should also be an additional
stochastic forcing term in (\ref{eq:coarse_grained_Strato}). This
term would balance the enhanced bare dissipation and restore fluctuation-dissipation
balance in the coarse-grained system \citet{OttingerBook}. A suitable
form of this term is not obvious and we do not include this term here
just as we did not include the stochastic forcing term $\grad\cdot\left(\sqrt{2\chi_{0}c}\,\M{\mathcal{W}}_{c}\right)$.

It is not possible to numerically solve (\ref{eq:limiting_Strato})
due to the presence of nontrivial dynamics at essentially \emph{all}
length scales, especially in the absence of bare diffusion. Our arguments
suggest that we can instead solve the coarse-grained equation (\ref{eq:coarse_grained_Strato}),
which has exactly the same form as (\ref{eq:limiting_Strato}), but
in which small scales are not resolved, and there is increased bare
dissipation. This is very easy to do in finite-volume numerical methods
for solving (\ref{eq:limiting_Strato}) by simply increasing the cell
volume and increasing the bare diffusion coefficient accordingly.
In the right panel of Fig. \ref{fig:SmoothDecay}, we show the result
of an Eulerian simulation performed with a four times coarser grid
than the middle panel. This is roughly equivalent to choosing $\delta=4\sigma$
and solving (\ref{eq:coarse_grained_Strato}). The value of $\chi_{0}$
is increased according to a discrete equivalent of (\ref{eq:chi_renormalization})
to account for the unresolved scales. This ensures that the effective
diffusion coefficient $\chi_{\text{eff}}$ is the same for all panels
of Fig. \ref{fig:SmoothDecay}, in agreement with (\ref{eq:filtered_c_Ito}).
Except at scales not resolved by the four-times coarser grid, the
right panel and the middle panel look similar visually, as confirmed
by an examination of the corresponding Fourier spectra. This suggests
that (\ref{eq:coarse_grained_Strato}) is indeed a good approximation
to the true dynamics of the spatially coarse-grained concentration.
A more quantitative comparison between the spatially smoothed $c{}_{\delta}$
and the conditional average $\bar{c}_{\delta}$ will be performed
in future studies.

We emphasize that the inclusion of the fluctuating term $-\V w_{\delta}\odot\grad c{}_{\delta}$
in (\ref{eq:coarse_grained_Strato}) is \emph{necessary} to obtain
the correct physical behavior, especially in two dimensions. In large
three dimensional systems, when the spatial coarse-graining is performed
at macroscopic scales $\delta\gg\sigma$, it has often been assumed
\citet{FluctHydroNonEq_Book} that one can approximate (\ref{eq:coarse_grained_Strato})
with the deterministic Fick's law (\ref{eq:dc_dt_mean}) and linearize
the fluctuations around the deterministic dynamics, as in (\ref{eq:limiting_linearized}).
To our knowledge there have been no precise mathematical arguments
to support this picture suggested by renormalization arguments \citet{DiffusionRenormalization_I}.
In two dimensions, linearization is certainly not appropriate due
to the logarithmic growth of the effective diffusion coefficient (\ref{eq:chi_SE})
with system size. Thin films may exhibit an intermediate behavior
depending on the scale of observation relative to the thickness of
the thin film \citet{ThinFilm_Smectic}.

\subsection{Irreversibility of Coarse-Graining}

In the coarse-grained dynamics (\ref{eq:coarse_grained_Strato}),
there is irreversible dissipation, $\D{\chi_{\delta}}>0$, even in
the absence of dissipation in the original dynamics. It is easy to
appreciate that elimination of degrees of freedom (coarse-graining)
is necessary in order to obtain dissipative (irreversible) dynamics
starting from a non-dissipative (reversible, even Hamiltonian) dynamics
\citet{OttingerBook}. Consider the specific example of diffusive
mixing illustrated in Fig. \ref{fig:DiffusiveInterface} in the absence
of bare diffusion, \textbf{$\chi_{0}=0$}. Since $\V u$ and $\V w$
are spatially-smooth velocity fields, advection by $\V u$ or $\V w$,
in the absence of bare diffusion, leads to behavior qualitatively
different from diffusion. Specifically, if the initial concentration
$c\left(\V r,0\right)$ has a sharp interface, this interface will
remain sharp at all times, even if it becomes very rough. This implies
that if $\chi_{0}=0$, in Fig. \ref{fig:DiffusiveInterface} one should
see only the red and blue colors present in the initial snapshot,
at \emph{all} times, in \emph{every} realization, instead of the spectrum
of colors actually seen in the figure.

\begin{figure*}[tbph]
\begin{centering}
\includegraphics[width=1\textwidth]{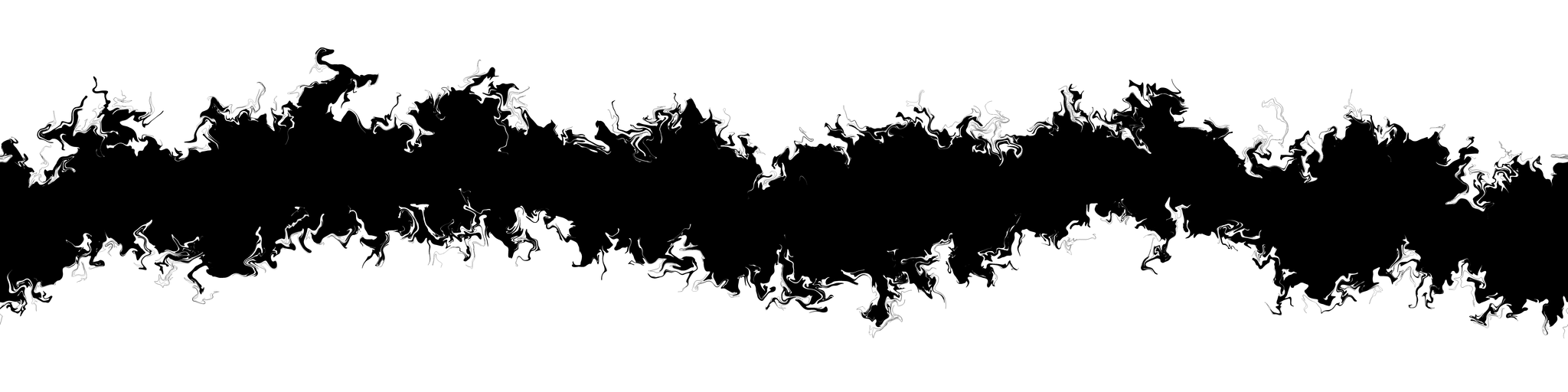}\vspace{0.1cm}
\par\end{centering}

\begin{centering}
\includegraphics[width=0.49\textwidth]{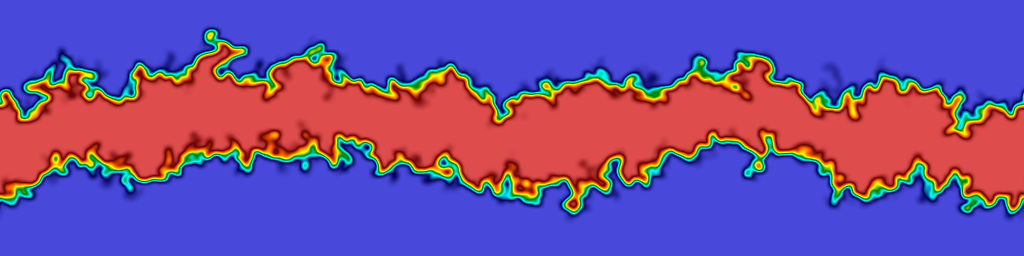}\hspace{0.1cm}\includegraphics[width=0.49\textwidth]{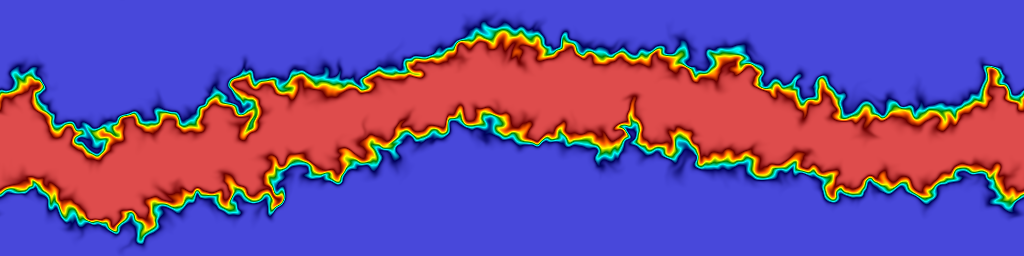}\vspace{0.1cm}
\par\end{centering}

\begin{centering}
\includegraphics[width=0.49\textwidth]{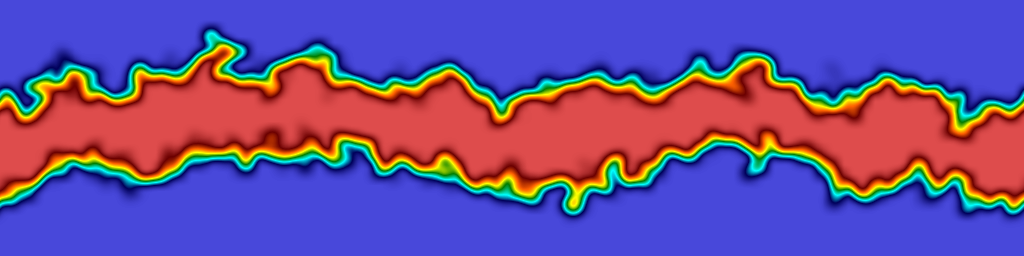}\hspace{0.1cm}\includegraphics[width=0.49\textwidth]{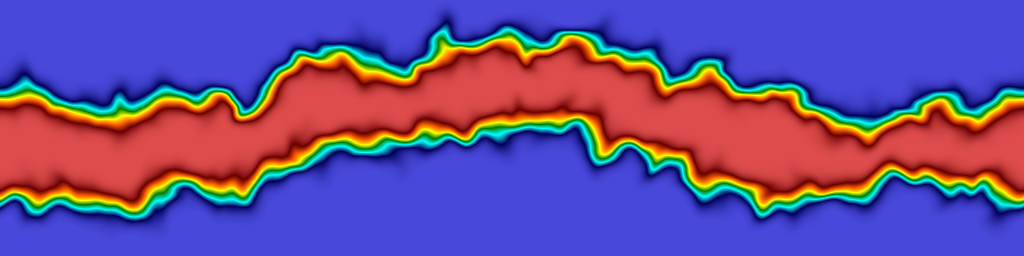}
\par\end{centering}

\caption{\label{fig:LagrangianEulerian}(\emph{Top panel}) A snapshot of the
concentration $c$ for the diffusive mixing process first shown in
Fig. \ref{fig:DiffusiveInterface}, here in the absence of bare diffusion,
$\chi_{0}=0$. The top and bottom interface are represented with about
half a million Lagrangian tracers each, and (\ref{eq:Lagrangian_Strato})
is solved for each tracer numerically. A Gaussian filter with standard
deviation $\sigma$ is used to filter the velocity field, and the
periodic domain has unit cell of shape $512\sigma\times128\sigma$.
The space between the two interfaces is colored black using image-processing
tools. (\emph{Two color panels on left}) The spatially-coarse grained
concentration $c{}_{\delta}$ obtained by blurring the top panel using
a Gaussian filter with standard deviation $\delta$, for $\delta=1.5\sigma$
(top left) and $\delta=3\sigma$ (bottom left). (\emph{Two color panels
on right}) An independent snapshot of the conditional average $\bar{c}_{\delta}$
at the same point in time as the panels on the left, obtained by solving
(\ref{eq:filtered_c_Strato}) with an Eulerian method using a grid
of $2048\times512$ finite-volume cells. A Gaussian filter of width
$\delta$ is used to filter the discrete velocity and the bare diffusion
$\chi_{0}$ is chosen such that $\chi_{\text{eff}}$ is the same as
in the Lagrangian simulations. In the top panel, $\delta=1.5\sigma$
(six grid cells) and $\chi_{\text{eff}}/\chi_{0}\approx9.6$, and
in the bottom panel $\delta=3\sigma$ and $\chi_{\text{eff}}/\chi_{0}\approx3.5$.
The same settings and random number sequence was used to generate
the random velocities for both panels on the right in order to facilitate
a direct comparison.}
\end{figure*}

We turn to our Lagrangian tracer algorithm for solving (\ref{eq:Lagrangian_Strato})
as a means to track the interface in Fig. \ref{fig:DiffusiveInterface}
without dissipation. For the particular example of diffusive mixing
starting from a sharp interface, in the absence of bare diffusion,
tracking the interface is sufficient to reconstruct the solution everywhere.
Specifically, $c=0$ on one side of the interface (topologically a
closed curve on the torus for a periodic system), and $c=1$ on the
other side. Therefore, we put a large number of Lagrangian tracers
on the flat interface at $t=0$, keeping the distance between neighboring
tracers much smaller than the molecular cutoff scale $\sigma$. We
then simulate a realization of the particles' trajectories to a later
time, connecting neighboring points with straight line segments to
obtain an approximation of the interface. In the top panel of Fig.
\ref{fig:LagrangianEulerian} we show the results of a Lagrangian
simulation of the mixing process first illustrated in Fig. \ref{fig:DiffusiveInterface}.
The top and bottom interface are tracked using tracers, and the concentration
in the space between the two interfaces is set to $c=1$ (black),
$c=0$ elsewhere (white).

\begin{figure}[tbph]
\begin{centering}
\includegraphics[width=0.49\textwidth]{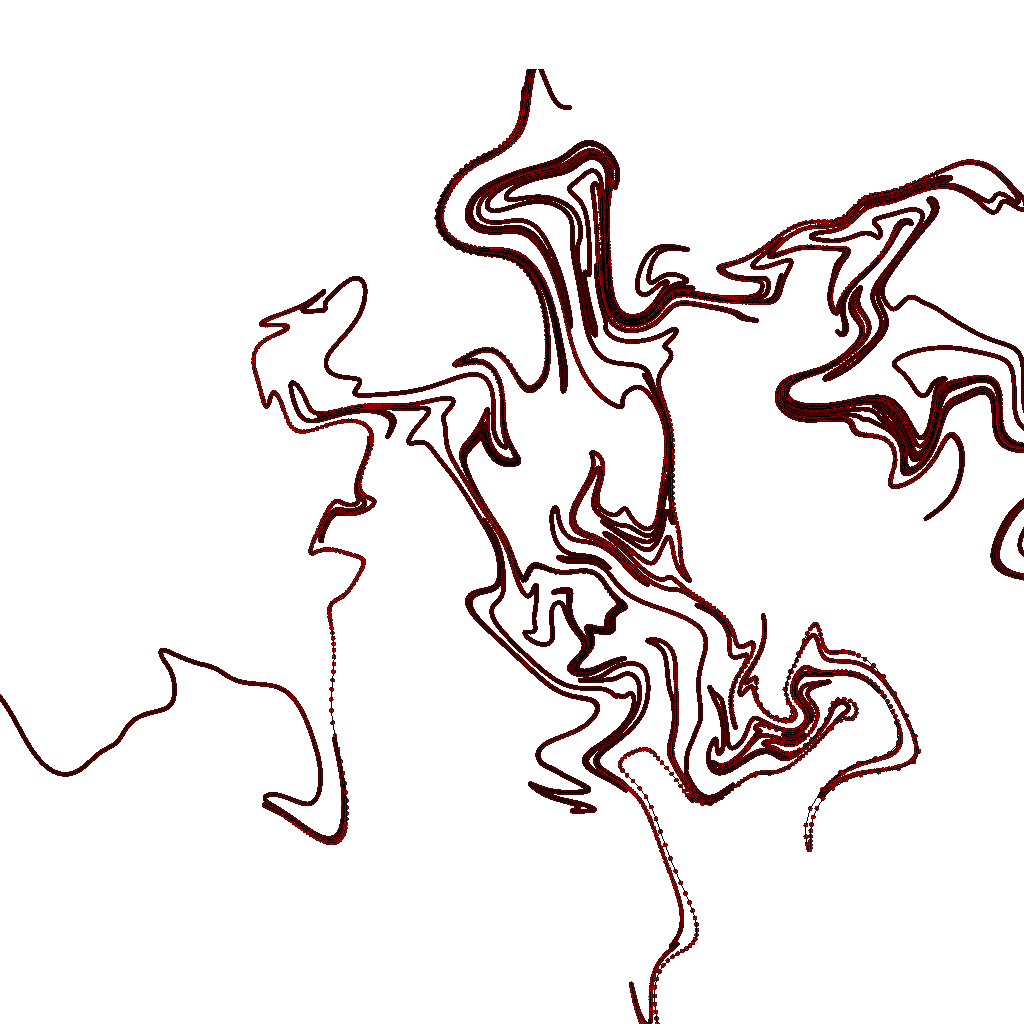}
\par\end{centering}

\caption{\textbf{\label{fig:LagrangianInterface}}A snapshot of a portion of
an initially straight line of Lagrangian tracers after some time.
The individual tracers and the straight line segments connecting them
are both shown. The length of the shown portion of the domain is about
$20\sigma$.}
\end{figure}

As illustrated in Fig. \ref{fig:LagrangianInterface}, an initially
straight line of tracers becomes quite contorted at later times, even
though topologically it remains a non-crossing curve at all times.
Asymptotically as $t\rightarrow\infty$ we expect that the line will
densely cover the plane (i.e., become a space-filling curve), in the
same way that simple diffusion would lead to uniform concentration
throughout the domain. Simulating the mixing process using a Lagrangian
algorithm would therefore require an unbounded increase in the number
of Lagrangian tracers with time in order to track the ever-increasing
level of fine-scale detail in the interface. Spatial coarse-graining
introduces effective bare diffusion and eliminates the fine-scale
details in the mixing front. In the two color panels on the left in
Fig. \ref{fig:LagrangianEulerian} we show the concentration field
$c{}_{\delta}=\M{\delta}\star c$ smoothed with a Gaussian filter
of width $\delta=1.5\sigma$ and $\delta=3\sigma$, now showing a
spectrum of colors due to the spatial averaging.

In the two color panels on the right in Fig. \ref{fig:LagrangianEulerian}
we show statistically-independent samples of the conditional average
$\bar{c}_{\delta}=\av c_{\tilde{\V w}}$ obtained by solving (\ref{eq:filtered_c_Strato})
using a finite-volume Eulerian algorithm. A slight modification of
the algorithm used to prepare Figs. \ref{fig:DiffusiveInterface},
\ref{fig:SmoothDecay} and \ref{fig:S_k_c} was implemented, in which
the discrete random advection velocity $\V w$ was filtered in Fourier
space with a Gaussian filter of width $\delta$ to obtain the $\V w_{\delta}$,
as discussed in more detail in Section \ref{sub:EulerianAlgorithm}.
The grid spacing was set to be smaller than $\delta/6$, which ensures
that discretization artifacts are quite small and the Eulerian code
can be directly compared to the very accurate Lagrangian code. The
value of the coarse-graining length was set to be $\delta>\sigma$,
and the bare diffusion coefficient $\D{\chi_{\delta}}>0$ was set
so that the effective diffusion coefficient $\chi_{\text{eff}}$ remained
the same as in the Lagrangian simulation. One can choose the balance
between bare diffusion and enhanced diffusion essentially arbitrarily
by choosing the length scale $\delta$ at which to truncate (filter)
the velocity spectrum, and for the top right color panel of the figure
we used the smallest value of $\delta=1.5\sigma$ that stabilized
the numerical method. The coarse-graining length $\delta=3\sigma$
is twice larger in the bottom right color panel than in the top right
color panel, and therefore there is enhanced bare diffusion (smoothing).
Visually the two color panels on the left and on the right in Fig.
\ref{fig:LagrangianEulerian} look quite similar. This indicates that
(\ref{eq:coarse_grained_Strato}) gives a good approximation to the
nonequilibrium fluctuations in the coarse-grained field $c{}_{\delta}$.

\subsection{A Paradigm for Diffusion}

Let us now summarize our discussion of spatial coarse-graining. We
start from the overdamped equation (\ref{eq:limiting_Strato}) as
the most accurate representation of diffusion, although itself an
approximation of the true molecular transport processes. The reference
molecular scale $\sigma$ and bare diffusion coefficient $\chi_{0}$
may in principle be extracted from comparisons to a more fundamental
model such as molecular dynamics, or from experimental observations.
In the end, the precise details of the dynamics at the molecular scale
do not matter, since at the larger scales they only enter through
a renormalized bare diffusion coefficient. In fact, the microscopic
equation (\ref{eq:limiting_Strato}) should never be solved directly.
Doing so numerically would require using a grid resolution smaller
than the molecular scale, and, in the case of no bare diffusion, would
require an infinite resolution due to the creation of every finer-scale
details in the solution even when the initial condition is smooth.
Instead, what one should really calculate is the spatially-coarse
grained $c{}_{\delta}=\M{\delta}\star c$, where $\delta\gg\sigma$
is a scale of observation.

In order to derive an approximation for the dynamics of $c{}_{\delta}$,
we started by splitting the spectrum of the velocity fluctuations
into a microscopic component $\tilde{\V w}$ containing the fluctuations
at scales below a mesoscopic length $\delta$, and the rest of the
spectrum extending all the way to the macroscopic scale. A rigorous
closed-form equation for the conditional average $\bar{c}_{\delta}=\av c_{\tilde{\V w}}$
is given by (\ref{eq:filtered_c_Strato}). A key result of our numerical
experiments illustrated in Fig. \ref{fig:LagrangianEulerian} is that
$c{}_{\delta}\approx\bar{c}_{\delta}$, more precisely, that (\ref{eq:filtered_c_Strato})
can be used to approximate the uknown dynamics of $c{}_{\delta}$.
We can express the relations between the different quantities by the
following diagram
\[
\begin{array}{ccc}
 &  & c{}_{\delta}=\M{\delta}\star c\\
 & \nearrow\\
c &  & \Updownarrow\\
 & \searrow\\
 &  & \bar{c}_{\delta}=\av c_{\tilde{\V w}}
\end{array}
\]
In the coarse-grained approximation (\ref{eq:coarse_grained_Strato}),
the bare diffusion coefficient is renormalized to take into account
the contribution of advection by the unrepresented (eliminated) velocity
scales. This makes solving this equation numerically a much simpler
task, since the enhanced bare diffusion dissipates small scale features
in the solution. 

The correspondence $c{}_{\delta}\Leftrightarrow\bar{c}_{\delta}$
is an approximation inspired by (\ref{eq:filtered_c_Strato}). Stochastic
homogenization theory \citet{TurbulenceClosures_Majda} can potentially
be used to justify (\ref{eq:coarse_grained_Strato}) for $\delta\gg\sigma$.
It is unlikely that there can be a rigorous justification for this
identification in the case when there is no separation of scales between
the mesoscopic and microscopic scales, i.e., when $\delta\sim\sigma$,
even though Fig. \ref{fig:LagrangianEulerian} shows a very good visual
agreement. In future work we will perform more detailed quantitative
comparisons in order to quantify the length and time scales at which
(\ref{eq:coarse_grained_Strato}) is a good approximation.

\section{\label{sec:Conclusions}Conclusions}

We presented a model of diffusion in liquids that captures in a simple
yet precise way the contribution that thermal velocity fluctuations
make to the transport of a passive tracer. The standard equations
of fluctuating hydrodynamics used to describe the effect of thermal
fluctuations on diffusion \citet{FluctHydroNonEq_Book} need to be
regularized below a cutoff molecular scale. We introduced this regularization
by filtering the fluctuating velocity field $\V v$ at a molecular
scale $\sigma$ in order to obtain a smooth (in both space and time)
velocity $\V u$ with which we advect the passive tracer. Under the
assumption of large separation of scales between the fast momentum
diffusion (collisional transport of momentum) and the slow mass diffusion,
i.e., large Schmidt number, we obtained an overdamped limiting equation
for the concentration. This equation is amenable to numerical simulations,
allowing us to simulate diffusive mixing even in the presence of infinite
separation of time scales between mass and momentum diffusion.

In the Stratonovich form the overdamped equation for the concentration
of passive tracers is a stochastic advection-diffusion equation in
which the thermal velocity fluctuations enter as a white-in-time random
advection field $\V w$ with spectrum given by a Green-Kubo formula.
For the case of Stokes flow the spectrum of $\V w$ is proportional
to a regularized Oseen tensor. In the Ito form of the overdamped equation,
there is an additional diffusive term with diffusion coefficient closely
related to the Stokes-Einstein prediction for the diffusion coefficient
of a sphere of radius $\sigma$ immersed in the fluid. This enhancement
of the diffusion over the bare diffusion is mathematically similar
to the well-known eddy diffusivity in turbulent transport. However,
its origin is very different physically since the random flow here
describes very low Reynolds number thermal fluctuations in the velocity.
Unlike previous derivations of the Stokes-Einstein law for diffusion
in liquids, our model makes no assumptions beyond that of a large
Schmidt number and gives a stochastic \emph{dynamical} description
of diffusion.

The sum of the bare and enhanced diffusion coefficients determines
the effective diffusion coefficient, which gives the rate of dissipation
in the ensemble mean. In each individual realization of the diffusion
process, however, bare diffusion and the random advection giving rise
to the enhanced diffusion behave rather differently because the random
advection is strictly non-dissipative. We showed that, from the perspective
of an initially excited mode (wavenumber), the advection by the random
velocity field effects apparent dissipation in the form of transfer
of power into other modes. The average rate of power dissipation is
found to be exactly the same as for simple diffusion with an equal
effective diffusion coefficient. However, the physical behavior of
each realization is very different from that predicted by Fick's deterministic
law of diffusion. Instead of each mode being independent of all other
modes as in simple (linear) diffusion, the random advection couples
all the modes and produces large-scale giant fluctuations in the concentration.
These are manifested in a power-law behavior of the spectrum of concentrations,
as predicted by linearized fluctuating hydrodynamics and observed
in recent experiments \citet{GiantFluctuations_Nature,GiantFluctuations_Cannell,FractalDiffusion_Microgravity}.

Here we studied coarse-graining based on a continuum rather than a
discrete microscopic model of diffusion. An alternative approach to
coarse-graining of diffusion is to consider purely discrete models
in which the coarse-grained variables are not smothed fields, as we
have done here, but rather, a collection of discrete variables associated
with coarse-graining volumes (cells) of length $\delta\gg\sigma$
\citet{DiscreteLLNS_Espanol,DiscreteDiffusion_Espanol}. The accuracy
of such finite-dimensional truncations can, in principle, be evaluated
by comparing them to particle simulations. As an alternative, one
can start from the more tractable continuum model (\ref{eq:limiting_Strato})
and think of finite-dimensional truncations as discretizations of
(\ref{eq:coarse_grained_Strato}). In the end, our numerical observations
suggest that at scales much larger than the molecular the behavior
of all models is similar, and can be described by a combination of
bare diffusion and advection by a thermally fluctuating velocity field.
Understanding this equivalence mathematically is a challenge common
to all dynamical coarse-graining endeavors.

In typical experiments, such as FRAP measurements of diffusion coefficients,
one observes the concentration spatially-coarse grained at scales
much larger than the molecular scale. We discussed how to perform
such spatial coarse-graining for the conditional ensemble average
over only the unresolved velocity fluctuations. This conditional mean
shows true dissipation in the form of a renormalized diffusion coefficient,
a remnant of the eliminated degrees of freedom. We observed numerically
that the equation for the conditional ensemble average is a good approximation
(closure) to the dynamics of individual realizations of the spatially
coarse-grained concentration. This means that, even in the absence
of bare diffusion, the coarse-grained concentration shows dissipative
behavior, as we confirmed using Lagrangian numerical simulations.
The renormalized diffusion coefficient in the coarse-grained equation
is nonzero even in the absence of bare diffusion, and, in fact, one
can set $\chi_{0}=0$ without affecting the behavior of the concentration
field at mesoscopic scales. The renormalized diffusion coefficient
is then controlled by the molecular scale $\sigma$ only, in agreement
with Stokes-Einstein's formula. Contrary to the standard renormalization
theory \citet{DiffusionRenormalization_I,DiffusionRenormalization_II}
which accounts for the contribution of thermal fluctuations as a perturbation
(correction) to the bare (molecular) diffusion coefficient, in our
model diffusion arises \emph{entirely} due to the velocity fluctuations
and it is not necessary to include an \emph{ad hoc} bare diffusion
term.

In the limit of infinite coarse-graining length scale, at least in
three dimensions, one expects to obtain the usual Fick's law of diffusion.
That is, we expect that at macroscopic scales one recovers the deterministic
diffusion equation (\ref{eq:dc_dt_mean}) not just for the ensemble
mean but also (as a law of large numbers) for each instance (realization)
of the mixing process. Understanding the precise relationship between
the macroscopic and microscopic dynamics is a challenge even for much
simpler models of diffusion such as the case of non-interacting Brownian
walkers \citet{LDT_Excluded,LDT_NonExcluded}; it therefore remains
an important future challenge to understand the precise relationship
between Fick's law, (\ref{eq:limiting_Ito}) (equivalently, (\ref{eq:limiting_Strato}))
and (\ref{eq:limiting_linearized}), in three dimensions. In two dimensions,
however, there is no macroscopic limit because the renormalized diffusion
coefficient grows logarithmically with system size. 

More importantly, in both two and three dimensions the behavior of
a diffusive mixing process cannot be described by Fick's law at mesoscopic
scales. One must include random advection by the mesoscopic scales
of the velocity fluctuations in order to reproduce not just the behavior
of the mean but also the giant fluctuations observed in individual
realizations (instances). The diffusion renormalization depends sensitively
on the spectrum of the velocity fluctuations, which is affected by
boundary conditions (confinement) \citet{Nanopore_Fluctuations,ThinFilm_Smectic,DiffusionRenormalization}.
The traditional Fick's diffusion constant is only meaningful under
special conditions which may not in fact be satisfied in many experiments
aimed to measure ``the'' diffusion coefficient. A length scale of
observation (coarse-graining) must be attached to the diffusion coefficient
value in order to make it a true ``material constant'' that can
be used in a predictive model of diffusive transport \citet{DiffusionRenormalization}.

Dismissing the effect of thermal fluctuations as ``weak'' is easy
with hand-waving estimates, but not easily justified upon an in-depth
analysis as we have performed here. We hope that our work will spur
interest in designing experiments that carefully examine diffusion
at a broad range of length scales. Existing experiments have been
able to measure concentration fluctuations across a wide range of
lenghtscales transverse to the gradient, but fluctuations are averaged
longitudinally over essentially macroscopic scales (thickness of the
sample) \citet{GiantFluctuations_Nature,GiantFluctuations_Cannell,FractalDiffusion_Microgravity}.
FRAP experiments routinely look at diffusion at micrometer scales,
however, we are not aware of any work that has even attempted to account
for the important effect of thermal fluctuations. It is perhaps not
surprising that diffusion coefficients in liquids are typically only
known to at most a couple of decimal places. The renormalization of
the diffusion coefficient by the velocity fluctuations depends on
the geometry of the sample and the initial excitation, and on factors
such as gravity and surface tension. Giant fluctuations are expected
to be more easily observed and measured in thin liquid films due to
the quasi-two dimensional geometry \citet{ThinFilm_Smectic,LiquidCrystalFilms}.

Recently, nonequilibrium fluctuations have been used as a way to measure
mass and thermophoretic diffusion coefficients more accurately \citet{SoretDiffusion_Croccolo}.
Our work is directly relevant to such efforts, especially when combined
with numerical methods to solve the resulting stochastic advection-diffusion
equations \citet{LLNS_Staggered,LowMachExplicit}. The simple model
we considered here is only applicable to self-diffusion or diffusion
of tracers in the dilute regime. Generalizing the model and in particular
the mode-elimination procedure to more realistic binary fluid mixtures
is an important future research direction. To our knowledge, there
have been no studies of the renormalization of diffusion by thermal
velocity fluctuations in ternary mixtures. In the future we will consider
extensions of our approach to multispecies liquid mixtures. Such extensions
are expected to lead to a better understanding of the physics of diffusion
in fluid mixtures, including a generalized Stokes-Einstein relation
for inter-diffusion coefficients in dilute multispecies solutions.

\rule[0.5ex]{0.75\columnwidth}{1pt}
\begin{acknowledgments}
We would like to acknowledge Florencio Balboa Usabiaga and Andreas
Klockner for their help in developing a GPU implementation of the
numerical methods, and Leslie Greengard for advice on the non-uniform
FFT algorithm. We are grateful to Alberto Vailati, Alejandro Garcia,
John Bell, Sascha Hilgenfeldt, Mike Cates and Ranojoy Adhikari for
their insightful comments.\emph{ }A. Donev was supported in part by
the National Science Foundation under grant DMS-1115341 and the Office
of Science of the U.S. Department of Energy through Early Career award
DE-SC0008271. T. Fai acknowledges the support of the DOE Computational
Science Graduate Fellowship, under grant number DE-FG02-97ER25308.
E. Vanden-Eijnden was supported by the DOE office of Advanced Scientific
Computing Research under grant DE-FG02-88ER25053, by the NSF under
grant DMS07-08140, and by the Office of Naval Research under grant
N00014-11-1-0345.
\end{acknowledgments}
\appendix

\section*{Appendix}

\section{\label{sec:ModeElimination}Mode Elimination Procedure}

In this appendix we consider the system of equations (\ref{eq:v_eq},\ref{eq:smoothing_u},\ref{eq:c_eq_original})
in the limit of infinite Schmidt number. For completeness, and because
gravity is known to strongly affect giant fluctuations \citet{GiantFluctConcentration_Sengers,GiantFluctuations_Cannell,GiantFluctuations_Theory,FractalDiffusion_Microgravity}
in actual experiments, we include here a buoyancy term in the velocity
(momentum) equation. This introduces a coupling of concentration back
into the velocity equation. It is convenient to eliminate the incompressibility
constraint by using a projection operator formalism to remove the
pressure from the fluctuating Navier-Stokes equation, 
\begin{align}
\rho\partial_{t}\V v= & \M{\mathcal{P}}\left[\eta\grad^{2}\V v+\grad\cdot\left(\sqrt{2\eta k_{B}T}\,\M{\mathcal{W}}\right)-\beta\rho c\,\V g\right]\label{eq:00}\\
\partial_{t}c= & -\V u\cdot\grad c+\chi_{0}\grad^{2}c+\grad\cdot\left(\sqrt{2\chi_{0}c}\,\M{\mathcal{W}}_{c}\right).\label{eq:00_c}
\end{align}
Here $\V u=\V{\sigma}\star\V v$ is the mollified version of $\V v$
defined in~\eqref{eq:smoothing_u}, $\beta$ is the solutal expansion
coefficient (assumed constant), and $\V g$ is the graviational acceleration;
$\M{\mathcal{P}}$ is the orthogonal projection onto the space of
divergence-free velocity fields, $\M{\mathcal{P}}=\M I-\M{\mathcal{G}}\left(\M{\mathcal{D}}\M{\mathcal{G}}\right)^{-1}\M{\mathcal{D}}$
in real space, where $\M{\mathcal{D}}\equiv\grad\cdot$ denotes the
divergence operator and $\M{\mathcal{G}}\equiv\grad$ the gradient
operator, with the appropriate boundary conditions taken into account.
With periodic boundaries we can express all operators in Fourier space
and $\widehat{\M{\mathcal{P}}}=\M I-k^{-2}(\V k\V k^{T})$, where
$\V k$ is the wave number. Note that our inclusion of the problematic
multiplicative-noise term $\grad\cdot\left(\sqrt{2\chi_{0}c}\,\M{\mathcal{W}}_{c}\right)$
is purely formal, as a precise interpretation of this term is missing.
For the purposes of this calculation we simply carry that term through
the calculation.

In this Appendix, we formally show that there exists a limiting dynamics
for $c$ as the bare Schmidt number $\Sc_{0}=\eta/\left(\rho\chi_{0}\right)\rightarrow\infty$
and $\chi_{0}\rightarrow0$ in such a way that 
\[
\chi_{0}^{2}\Sc_{0}\sim\chi_{0}\eta=\mbox{const},
\]
which is consistent with the scaling of the diffusion coefficient
with viscosity predicted by the Stokes-Einstein relation (\ref{eq:chi_SE}).
To this end, consider a family of equations in which the coefficients
are rescaled as 
\begin{equation}
\eta\mapsto\epsilon^{-1}\eta,\qquad\chi_{0}\mapsto\epsilon\chi_{0}.\label{eq:18}
\end{equation}
This rescaling preserves the product $\chi_{0}\eta$ but implies that
$\ensuremath{\Sc_{0}\mapsto\epsilon^{-2}\Sc_{0}}$, so that the rescaled
$\Sc_{0}\to\infty$ as $\epsilon\rightarrow0$. For $\epsilon=1$
we get the original dynamics (\ref{eq:00},\ref{eq:00_c}), and as
$\epsilon\rightarrow0$ we get the dynamics in the limit of infinite
Schmidt number. If the separation of time scales in the original dynamics
is sufficiently strong the limiting dynamics $\epsilon\rightarrow0$
is a good proxy for the real dynamics $\epsilon=1$. This assumption
of separation of time scales has to be verified \emph{a posteriori},
after the limiting dynamics is obtained; specifically, the actual
dimensionless number of interest is not the bare $\Sc_{0}$ but rather
the effective $\Sc=\eta/\left(\rho\chi_{\text{eff}}\right)\to\infty$.

Writing~\eqref{eq:00} in terms of the rescaled coefficients~\eqref{eq:18}
and rescaling time as $t\mapsto\epsilon^{-1}t$, we arrive at the
rescaled system 
\begin{align}
\partial_{t}\breve{\V v}= & \M{\mathcal{P}}\left[\epsilon^{-2}\eta\rho^{-1}\grad^{2}\breve{\V v}+\grad\cdot\left(\sqrt{2\epsilon^{-2}\eta\rho^{-2}\, k_{B}T}\,\M{\mathcal{W}}(t)\right)-\epsilon^{-1}\beta\breve{c}\,\V g\right]\label{eq:000}\\
\partial_{t}\breve{c}= & -\epsilon^{-1}\breve{\V u}\cdot\grad\breve{c}+\chi_{0}\grad^{2}\breve{c}+\grad\cdot\left(\sqrt{2\chi_{0}\breve{c}}\,\M{\mathcal{W}}_{c}(t)\right).\label{eq:000_c}
\end{align}
These equations define a Markov process with generator $\M L=\M L_{0}+\epsilon^{-1}\M L_{1}+\epsilon^{-2}\M L_{2}$,
where 
\begin{equation}
\begin{aligned}\M L_{0}F & =\chi_{0}\int d\V r\ \grad^{2}c(\V r)\frac{\delta F}{\delta c(\V r)}+\chi_{0}\int d\V r\ c(\V r)\grad^{2}\frac{\delta^{2}F}{\delta c(\V r)^{2}}\\
\M L_{1}F & =-\int d\V r\ \V u(\V r)\cdot\grad c(\V r)\frac{\delta F}{\delta c(\V r)}+\beta\int d\V r\ \M{\mathcal{P}}(c(\V r)\V g)\cdot\frac{\delta F}{\delta\V v(\V r)}\\
\M L_{2}F & =\eta\rho^{-1}\int d\V r\ \M{\mathcal{P}}\grad^{2}\V v\cdot\frac{\delta F}{\delta\V v(\V r)}+\eta\rho^{-2}k_{B}T\;\int d\V r\M{\mathcal{P}}\grad\grad:\frac{\delta^{2}F}{\delta\V v(\V r)^{2}},
\end{aligned}
\label{eq:4}
\end{equation}
and $\delta F/\delta c(\V r)$ denotes the functional derivative of
the functional $F\equiv F[c,\V v]$ with respect to the field $c(\V r)$
and similarly for $\delta F/\delta\V v(\V r)$. Note that the operator
$\M L_{2}$ is the generator of the so-called virtual fast process,
which is nothing more than the equation for the fast (i.e., the fluctuating)
velocity component written in its natural time scale $\tau=t/\epsilon^{2}$,
\begin{equation}
\partial_{\tau}\T{\V v}=\M{\mathcal{P}}\left[\eta\rho^{-1}\grad^{2}\T{\V v}+\grad\cdot\left(\sqrt{2\eta\rho^{-2}\, k_{B}T}\,\M{\mathcal{W}}(\tau)\right)\right].\label{eq:virtual}
\end{equation}
This process describes the dynamics of equilibrium fluctuations of
velocity in the Stokes regime. The dynamics is time-reversible with
respect to a Gaussian Gibbs-Boltzmann equilibrium distribution which,
for a periodic system, can be formally written as \citet{DFDB} 
\begin{equation}
P_{\text{eq}}\left(\T{\V v}\right)=Z^{-1}\exp\left[-\frac{\int d\V r\,\rho\tilde{v}^{2}}{2k_{B}T}\right]\V{\delta}\left(\int d\V r\,\rho\T{\V v}\right)\delta\left(\grad\cdot\T{\V v}\right).\label{eq:P_eq_v}
\end{equation}

The mathematical asymptotic expansion techniques we employ follow
the procedure introduced in \citet{Averaging_Khasminskii,Averaging_Kurtz,ModeElimination_Papanicolaou}
(for a review see also the book \citet{AveragingHomogenization}).
Denote by $\left(\breve{c}(\V r,t),\,\breve{\V v}(\V r,t)\right)$
the solution to (\ref{eq:000},\ref{eq:000_c}) for the initial conditions
$\left(\breve{c}(\V r,0),\,\breve{\V v}(\V r,0)\right)=\left(c(\V r),\,\V v(\V r)\right)$
and consider
\begin{equation}
\langle F[\breve{c}(\cdot,t)]\rangle\equiv G[c(\cdot),\V v(\cdot),t],\label{eq:6}
\end{equation}
where the expectation $\langle\cdot\rangle$ is taken over the realizations
of the noise terms $\M{\mathcal{W}}(t)$ and $\M{\mathcal{W}}_{c}(t)$.
This expectation defines a time-dependent functional $G$ of the initial
conditions which satisfies the (functional) backward Kolmogorov equation
\begin{equation}
\partial_{t}G=\M L_{0}G+\epsilon^{-1}\M L_{1}G+\epsilon^{-2}\M L_{2}G,\qquad G|_{t=0}=F.\label{eq:7}
\end{equation}
We wish to take the limit as $\epsilon\to0$ of this equation. To
this end, formally expand $G$ as 
\begin{equation}
G=G_{0}+\epsilon G_{1}+\epsilon^{2}G_{2}+\cdots\label{eq:8}
\end{equation}
and insert this expression in~\eqref{eq:7}, and collect terms of
increasing power in $\epsilon$. This gives the hierarchy 
\begin{equation}
\begin{aligned}\M L_{2}G_{0} & =0,\\
\M L_{2}G_{1} & =-\M L_{1}G_{0},\\
\M L_{2}G_{2} & =\partial_{t}G_{0}-\M L_{0}G_{0}-\M L_{1}G_{1},\\
 & \ \,\vdots
\end{aligned}
\label{eq:9}
\end{equation}

The first equation in~\eqref{eq:9} indicates that $G_{0}$ is a
functional of $c(\V r)$ alone, rather than $c(\V r)$ and $\V v(\V r)$,
i.e. 
\begin{equation}
G_{0}\equiv G_{0}[c].\label{eq:5}
\end{equation}
The second equation in~\eqref{eq:9} requires a solvability condition,
namely that its right hand side be in the range of the operator $\M L_{2}$.
Equivalently, the expectation of any term involving $\V v(\V r)$
in this right hand side with respect to the invariant measure of the
virtual fast process $\T{\V v}(\V r,t)$ given by \eqref{eq:P_eq_v}
must be zero. Denoting this expectation by $\av f_{\V v}=\int f\left(\T{\V v}\right)\, P_{\text{eq}}\left(\T{\V v}\right)\,\mathcal{D}\T{\V v}$,
where the integral is a formal functional integral, the solvability
condition can be written as 
\begin{equation}
0=\av{\M L_{1}G_{0}}_{\V v}\equiv-\int d\V r\av{\V u(\V r)}_{\V v}\cdot\grad c(\V r)\frac{\delta G_{0}}{\delta c(\V r)}\label{eq:10}
\end{equation}
where we used the fact that $G_{0}$ is a functional of $c$ alone
from~\eqref{eq:5}. This solvability condition is automatically satisfied
since $\av{\V v(\V r)}_{\V v}=0$. As a result, the second equation
in~\eqref{eq:9} can be solved in $G_{1}$ to obtain 
\begin{equation}
G_{1}=-\M L_{2}^{-1}\M L_{1}G_{0}\label{eq:12}
\end{equation}
where $\M L_{2}^{-1}$ denotes the pseudo-inverse of $\M L_{2}$.
The third equation in~\eqref{eq:9} also requires a solvability condition,
which reads 
\begin{equation}
\begin{aligned}\partial_{t}G_{0} & =\av{\M L_{0}G_{0}}_{\V v}+\av{\M L_{1}G_{1}}_{\!\!\V v}\\
 & =\M L_{0}G_{0}-\left<\M L_{1}\M L_{2}^{-1}\M L_{1}G_{0}\right>_{\!\!\V v}
\end{aligned}
\label{eq:14}
\end{equation}
where we used $\av{\M L_{0}G_{0}}_{\V v}=\M L_{0}G_{0}$ as well as~\eqref{eq:12}
to get the second equality.

To write the second term on the right hand side of~\eqref{eq:14}
explicitly notice that 
\begin{equation}
\M L_{1}G_{0}=-\int d\V r\ \V u(\V r)\cdot\grad c(\V r)\frac{\delta G_{0}}{\delta c(\V r)}\label{eq:2a2a}
\end{equation}
Since this operator is linear in $\V u(\V r)=\M{\sigma}\star\V v(\V r)$,
to compute the action of $\M L_{2}^{-1}$ on it, we can use 
\begin{equation}
\begin{aligned}\M L_{2}^{-1}\V u(\V r) & =\M L_{2}^{-1}\M{\sigma}\star\V v(\V r)=\M{\sigma}\star\M L_{2}^{-1}\V v(\V r)\\
 & =-\M{\sigma}\star\int_{0}^{\infty}d\tau\, e^{\tau\M L_{2}}\V v(\V r)=-\M{\sigma}\star\int_{0}^{\infty}d\tau\,\langle\T{\V v}(\V r,\tau)\rangle,
\end{aligned}
\label{eq:1a1a}
\end{equation}
where $\T{\V v}(\V r,\tau)$ denotes the solution to~\eqref{eq:virtual}
for the initial condition $\T{\V v}(\V r,0)=\V v(\V r)$ and the expectation
is the same as in~\eqref{eq:6}. From~\eqref{eq:virtual}, this
solution can be formally expressed as 
\begin{equation}
\begin{aligned}\T{\V v}(\V r,\tau) & =\exp\left(-\tau\eta\rho^{-1}\M{\mathcal{L}}\right)\V v(\V r)\\
 & +\int_{0}^{\tau}d\tau^{\prime}\ \exp\left(-(\tau-\tau^{\prime})\eta\rho^{-1}\M{\mathcal{L}}\right)\grad\cdot\left(\sqrt{2\eta\rho^{-2}\, k_{B}T}\,\M{\mathcal{W}}(\tau^{\prime})\right),
\end{aligned}
\label{eq:19}
\end{equation}
where $\M{\mathcal{L}}=-\M{\mathcal{P}}\grad^{2}$ is the Stokes operator.
The second term is linear in $\M{\mathcal{W}}$ and therefore has
zero average and does not contribute to the expectation in~\eqref{eq:1a1a}
(i.e., it is a martingale). As a result, combining~\eqref{eq:1a1a}
and~\eqref{eq:19} we conclude that 
\begin{equation}
\M L_{2}^{-1}\V u(\V r)=-\rho\eta^{-1}\left(\M G_{\sigma}\star\V v\right)(\V r)\equiv-\rho\eta^{-1}\int d\V r^{\prime}\,\M G_{\sigma}(\V r,\V r^{\prime})\V v(\V r^{\prime})\label{eq:1b1b}
\end{equation}
where $\M G_{\sigma}=\M{\sigma}\star\M G$ and $\M G(\V r,\V r^{\prime})$
is the Green's function for Stokes flow. More explicitly, $\M G_{\sigma}\star\V v\equiv\M{\sigma}\star\M{\mathcal{L}}^{-1}\V v$
is a shorthand notation for the smoothed solution of the Stokes equation
with unit viscosity: $\V w_{\sigma}=\M G_{\sigma}\star\V v$ if $\V w_{\sigma}=\M{\sigma}\star\V w$
and $\V w$ solves 
\begin{equation}
\grad\pi=\grad^{2}\V w+\V v,\qquad\grad\cdot\V w=0.\label{eq:22}
\end{equation}

Using~\eqref{eq:1b1b} and the obvious identity 
\begin{equation}
\frac{\delta c(\V r^{\prime})}{\delta c(\V r)}=\delta(\V r-\V r^{\prime}),\label{eq:20}
\end{equation}
we see that the second term on the right hand side of~\eqref{eq:14}
can be written as 
\begin{equation}
\begin{aligned} & -\left<\M L_{1}\M L_{2}^{-1}\M L_{1}G_{0}\right>_{\!\!\V v}\\
 & =\rho\eta^{-1}\int d\V rd\V r^{\prime}\ \grad c(\V r)\cdot\frac{\delta}{\delta c(\V r)}\left(\left<(\M{\sigma}\star\V v(\V r))\otimes(\M G_{\sigma}\star\V v(\V r^{\prime}))\right>_{\V v}\cdot\grad^{\prime}c(\V r^{\prime})\frac{\delta G_{0}}{\delta c(\V r^{\prime})}\right)\\
 & -\beta\rho\eta^{-1}\int d\V rd\V r^{\prime}\ \M{\mathcal{P}}(c\V (\V r)\V g)\cdot\left<\frac{\delta}{\delta\V v(\V r)}(\M G_{\sigma}\star\V v(\V r^{\prime}))\right>_{\V v}\cdot\grad^{\prime}c(\V r^{\prime})\frac{\delta G_{0}}{\delta c(\V r^{\prime})}\\
 & =\rho\eta^{-1}\int d\V rd\V r^{\prime}\ \grad c(\V r)\cdot\left<(\M{\sigma}\star\V v(\V r))\otimes(\M G_{\sigma}\star\V v(\V r^{\prime}))\right>_{\V v}\cdot\left(\grad^{\prime}c(\V r^{\prime})\frac{\delta^{2}G_{0}}{\delta c(\V r)\delta c(\V r^{\prime})}-\grad^{\prime}\delta(\V r^{\prime}-\V r)\frac{\delta G_{0}}{\delta c(\V r^{\prime})}\right)\\
 & -\beta\rho\eta^{-1}\int d\V rd\V r^{\prime}\ \M{\mathcal{P}}(c\V (\V r)\V g)\cdot\left<\frac{\delta}{\delta\V v(\V r)}(\M G_{\sigma}\star\V v(\V r^{\prime}))\right>_{\V v}\cdot\grad^{\prime}c(\V r^{\prime})\frac{\delta G_{0}}{\delta c(\V r^{\prime})},
\end{aligned}
\label{eq:15a}
\end{equation}
where $\grad^{\prime}$ denotes the gradient operator with respect
to $\V r^{\prime}$. The first equality will be useful to write the
limiting equation for $c$ in Stratonovich's form, and the second
one in Ito's form.

To proceed further we need to explicitly perform the averages over
the equilibrium distribution of the fast virtual process, 
\begin{equation}
\rho\eta^{-1}\begin{aligned}\left<(\M{\sigma}\star\V v(\V r))\otimes(\M G_{\sigma}\star\V v(\V r^{\prime}))\right>_{\V v} & =\tfrac{1}{2}\M{\mathcal{R}}(\V r,\V r^{\prime}),\end{aligned}
\label{eq:1c1c}
\end{equation}
where $\M{\mathcal{R}}(\V r,\V r^{\prime})$ is the tensor defined
in~\eqref{eq:R_r}, and (using $\delta\V v(\V r^{\prime})/\delta\V v(\V r)=\M{\delta}(\V r-\V r^{\prime})$)
\begin{equation}
\begin{aligned} & \int d\V r\ \M{\mathcal{P}}(c\V (\V r)\V g)\cdot\left<\frac{\delta}{\delta\V v(\V r)}\left(\M G_{\sigma}\star\V v(\V r^{\prime})\right)\right>_{\V v}\\
 & =\int d\V rd\V r^{\prime\prime}\ \M{\mathcal{P}}(c\V (\V r)\V g)\cdot\left<\frac{\delta}{\delta\V v(\V r)}\left(\M G_{\sigma}(\V r^{\prime},\V r^{\prime\prime})\V v(\V r^{\prime\prime})\right)\right>_{\V v}\\
 & =\int d\V rd\V r^{\prime\prime}\ \M{\mathcal{P}}(c\V (\V r)\V g)\left(\M G_{\sigma}(\V r^{\prime},\V r^{\prime\prime})\M{\delta}(\V r-\V r^{\prime\prime})\right)\\
 & =\int d\V r^{\prime\prime}\ \M G_{\sigma}(\V r^{\prime},\V r^{\prime\prime})\M{\mathcal{P}}(c\V (\V r^{\prime\prime})\V g)\\
 & =(\M G_{\sigma}\star\M{\mathcal{P}}c)(\V r^{\prime})\,\V g.
\end{aligned}
\label{eq:2cc2cc}
\end{equation}
Inserting~\eqref{eq:1c1c} and~\eqref{eq:2cc2cc} in~\eqref{eq:15a},
performing an integration by part and using the property that $\grad\cdot\M{\mathcal{R}}(\V r,\V r^{\prime})=0$
(which follows from $\grad\cdot\V u=0$) and $\M G_{\sigma}\star\M{\mathcal{P}}=\M G_{\sigma}$
(which follows from $\M{\mathcal{P}}\M{\mathcal{L}}^{-1}=\M{\mathcal{L}}^{-1}\M{\mathcal{P}}=\M{\mathcal{L}}^{-1}$)
we finally obtain 
\begin{equation}
\begin{aligned} & -\left<\M L_{1}\M L_{2}^{-1}\M L_{1}G_{0}\right>_{\!\!\V v}\\
 & =\tfrac{1}{2}\int d\V rd\V r^{\prime}\ \grad c(\V r)\cdot\frac{\delta}{\delta c(\V r)}\left(\M{\mathcal{R}}(\V r,\V r^{\prime})\cdot\grad^{\prime}c(\V r^{\prime})\frac{\delta G_{0}}{\delta c(\V r^{\prime})}\right)\\
 & \quad+\beta\rho\eta^{-1}\int d\V r\ \left(\M G_{\sigma}\star c\right)\!(\V r)\V g\cdot\grad c(\V r)\frac{\delta G_{0}}{\delta c(\V r)}\\
 & =\tfrac{1}{2}\int d\V rd\V r^{\prime}\ \grad c(\V r)\cdot\M{\mathcal{R}}(\V r,\V r^{\prime})\cdot\grad^{\prime}c(\V r^{\prime})\frac{\delta^{2}G_{0}}{\delta c(\V r)\delta c(\V r^{\prime})}\\
 & \quad+\int d\V r\ \grad\cdot\left(\M{\chi}(\V r)\grad c(\V r)\right)\frac{\delta G_{0}}{\delta c(\V r)}\\
 & \quad+\beta\rho\eta^{-1}\int d\V r\ \left(\M G_{\sigma}\star c\right)\!(\V r)\V g\cdot\grad c(\V r)\frac{\delta G_{0}}{\delta c(\V r)},
\end{aligned}
\label{eq:15}
\end{equation}
where we recall $\M{\chi}(\V r)=\tfrac{1}{2}\M{\mathcal{R}}(\V r,\V r)$.

Inserting \eqref{eq:15} in \eqref{eq:14} gives the explicit form
of the limiting equation for $G_{0}=\lim_{\epsilon\to0}G$. This equation
is a backward Kolmogorov equation from which the limiting stochastic
differential equation for $c$ as $\epsilon\to0$ can be read. The
second-order functional derivative (with respect to $c(\V r)$) written
as in the first form of the right hand side of \eqref{eq:15} gives
this equation in Stratonovich's interpretation, while the second form
of the right hand side of \eqref{eq:15} gives it in Ito's interpretation:
\begin{equation}
\begin{aligned}dc & =-\sum_{k}\V{\phi}_{k}\cdot\grad c\ \circ d\mathcal{B}_{k}\\
 & \quad+\chi_{0}\grad^{2}c\ dt+\grad\cdot\left(\sqrt{2\chi_{0}c}\, d\M{\mathcal{B}}_{c}\right)\\
 & \quad+\beta\rho\eta^{-1}\left(\M G_{\sigma}\star c\right)\V g\cdot\grad c\, dt\\
 & =-\sum_{k}\V{\phi}_{k}\cdot\grad c\ d\mathcal{B}_{k}+\grad\cdot\left(\M{\chi}\grad c\right)dt\\
 & \quad+\chi_{0}\grad^{2}c\ dt+\grad\cdot\left(\sqrt{2\chi_{0}c}\, d\M{\mathcal{B}}_{c}\right)\\
 & \quad+\beta\rho\eta^{-1}\left(\M G_{\sigma}\star c\right)\V g\cdot\grad c\, dt
\end{aligned}
\label{eq:17}
\end{equation}
where $\M{\mathcal{B}}_{c}\left(\V r,t\right)$ is the Brownian sheet
process such that we formally have $d\M{\mathcal{B}}_{c}/dt=\M{\mathcal{W}}_{c}$.
Note that the derivation above, while formal, leaves no ambiguity
in terms of the interpretation of the first term at the right hand
sides of~\eqref{eq:17}.

If we set $\V g=0$ in this equation, \eqref{eq:17} reduces to \eqref{eq:limiting_Strato}
and \eqref{eq:limiting_Ito}. Also note that the gravity does not
affect the effective diffusion coefficient in the limit of infinite
Schmidt number, as it may for finite Schmidt numbers \citet{ExtraDiffusion_Vailati}.
However, the spectrum of the fluctuations in concentration at small
wavenumbers is strongly affected by buoyancy effects, and \eqref{eq:17}
is the nonlinear generalization of the existing linearized fluctuating
hydrodynamic theory for this effect \citet{GiantFluctConcentration_Sengers,GiantFluctuations_Cannell,GiantFluctuations_Theory,FractalDiffusion_Microgravity,FluctHydroNonEq_Book}.
Note that because \eqref{eq:17} is nonlinear, it is no longer possible
to take expectation values and write a closed equation for ensemble
averages, as it was in the purely linear case.

\section{\label{sec:TemporalScheme}Multiscale Integrators}

In this Appendix we describe the multiscale algorithms \citet{HMM_Stochastic,SeamlessMultiscale}
used to numerically solve the limiting Eulerian (\ref{eq:limiting_Strato})
and Lagrangian (\ref{eq:Lagrangian_Strato}) equations. These methods
rely on being able to solve the steady Stokes equations with random
forcing, more precisely, to generate a random velocity field $\V w\left(\V r\right)$
with spatial covariance
\[
\av{\V w\left(\V r_{1}\right)\otimes\V w\left(\V r_{2}\right)}=\frac{2k_{B}T}{\eta}\int\M{\sigma}\left(\V r_{1},\V r^{\prime}\right)\M G\left(\V r^{\prime},\V r^{\prime\prime}\right)\M{\sigma}^{T}\left(\V r_{2},\V r^{\prime\prime}\right)d\V r^{\prime}d\V r^{\prime\prime}.
\]
For the simulations described here we rely on periodic boundary conditions,
which means that the Fourier basis diagonalizes the Stokes operator
and therefore one can use the Fast Fourier Transform (FFT) to efficiently
solve the steady Stokes equations. This has helped us to implement
both the Eulerian and the Lagrangian algorithm on Graphics Processing
Units (GPUs), which has enabled simulations with as many as 16 million
degrees of freedom. While both algorithms and our codes work in either
two or three dimensions, in order to be able to study power law behavior
over many decades we focus in this work on two dimensional systems
(for $d=2$ we use up to $2048^{2}$ grid cells or wave-indices, but
for $d=3$ we are presently limited to at most $256^{3}$ grids due
to memory requirements).

In principle one can use either the Ito or Stratonovich forms of the
limiting dynamics. The only difference is in the temporal integrator,
namely, Ito equations can be integrated with the Euler-Maruyama (one-step)
scheme, while Stratonovich equations require the Euler-Heun (predictor-corrector)
scheme \citet{EulerHeun}. Here we use the Stratonovich form of the
equations because this ensures discrete fluctuation-dissipation balance
between the random advection (fluctuation) and the effective diffusion
(dissipation). For periodic boundaries, the Ito drift term $\left[\partial_{\V q}\cdot\M{\chi}\left(\V q\right)\right]dt$
in the Lagrangian equation (\ref{eq:Lagrangian_Strato}) vanishes,
and there is no difference between the different stochastic interpretations.

\subsection{\label{sub:EulerianAlgorithm}Eulerian Algorithm}

For completeness, we include a gravitational buoyancy term in the
velocity equation and present an algorithm for solving the limiting
equation \eqref{eq:17}, which includes the effect of gravity. Additional
background on the types of spatio-temporal integrations used can be
found in previous works by some of us \citet{LLNS_Staggered,LowMachExplicit,DFDB,StokesKrylov};
here we only sketch the basic features. We do not include the term
$\grad\cdot\left(\sqrt{2\chi_{0}c}\,\M{\mathcal{W}}_{c}\right)$ since
properly discretizing this multiplicative noise is nontrivial, and
largely irrelevant when studying nonequilibrium fluctuations. 

The overdamped Eulerian dynamics can be efficiently simulated using
the following Euler-Heun predictor-corrector temporal algorithm with
time step size $\D t$, which updates the concentration from time
step $n$ to time step $n+1$ (denoted here by superscript):
\begin{enumerate}
\item Generate a random advection velocity by solving the steady Stokes
equation with random forcing,
\begin{align*}
\grad\pi^{n} & =\eta\left(\grad^{2}\V v^{n}\right)+\grad\cdot\left(\sqrt{\frac{2\eta\, k_{B}T}{\D t\D V}}\,\M W^{n}\right)-\rho\beta c^{n}\V g\\
\grad\cdot\V v^{n} & =0,
\end{align*}
and compute $\V u^{n}=\M{\sigma}\star\V v^{n}$ by filtering. Here
$\M W^{n}$ are a collection of Gaussian random variates generated
independently at each time step, and $\D V$ is the volume of each
grid cell.
\item Do a predictor step for (\ref{eq:limiting_Strato}) by solving for
$\tilde{c}^{n+1}$, 
\[
\frac{\tilde{c}^{n+1}-c^{n}}{\D t}=-\V u^{n}\cdot\grad c^{n}+\chi_{0}\grad^{2}\left(\frac{c^{n}+\tilde{c}^{n+1}}{2}\right).
\]

\item If gravity is zero, set $\V u^{n+\frac{1}{2}}=\V u^{n}$, otherwise,
solve the steady Stokes equation
\begin{align*}
\grad\pi^{n+\frac{1}{2}} & =\eta\left(\grad^{2}\V v^{n+\frac{1}{2}}\right)+\grad\cdot\left(\sqrt{\frac{2\eta\, k_{B}T}{\D t\D V}}\,\M W^{n}\right)-\rho\beta\left(\frac{c^{n}+\tilde{c}^{n+1}}{2}\right)\V g\\
\grad\cdot\V v^{n+\frac{1}{2}} & =0,
\end{align*}
and compute $\V u^{n+\frac{1}{2}}=\M{\sigma}\star\V v^{n+\frac{1}{2}}$.
Note that the same random stress is used here as in the predictor.
\item Take a corrector step for concentration to compute $c^{n+1}$,
\[
\frac{c^{n+1}-c^{n}}{\D t}=-\V u^{n+\frac{1}{2}}\cdot\grad\left(\frac{c^{n}+\tilde{c}^{n+1}}{2}\right)+\chi_{0}\grad^{2}\left(\frac{c^{n}+c^{n+1}}{2}\right).
\]

\end{enumerate}
This scheme can be shown to be a weakly first-order accurate temporal
integrator for (\ref{eq:limiting_Strato}); it is a weakly second-order
method for the linearized equations (\ref{eq:limiting_linearized})
with gravity included \citet{DFDB}. Note that advection is treated
explicitly. The key to obtaining the correct diffusion enhancement
is the fact that the average of $c^{n}$ and $\tilde{c}^{n+1}$ is
used to evaluate the advective fluxes in the corrector step. The bare
diffusive fluxes can be obtained via any consistent temporal discretization;
here we use the Crank-Nicolson or implicit midpoint rule, but an explicit
midpoint rule (as used for the advective fluxes) can also be used
since the main stability limitation on the time step comes from the
advective Courant number.

We discretize the continuum equations in space using a staggered finite-volume
fluctuating hydrodynamics solver \citet{LLNS_Staggered} and use an
iterative Krylov linear solver \citet{StokesKrylov,NonProjection_Griffith}
to solve the steady Stokes equations. In this Eulerian algorithm the
difficulty is in discretizing advection, as is well-known from turbulence
modeling. Because of the transfer of power from the coarse to the
fine scales, advection creates fine-scale features in the solution
that cannot be represented on the fixed Eulerian grid. This leads
to well-known Gibbs instability, and requires introducing some form
of dissipation at the larger wavenumbers. If there is sufficient bare
diffusion present to smooth the solution at the scale of the grid,
then one can use a strictly non-dissipative discrete advection operator
\citet{LLNS_Staggered}. This strictly non-dissipative centered advection
maintain discrete fluctuation-dissipation balance \citet{DFDB}, and
for this reason we have used it for the simulations reported here.
If there is insufficient bare diffusion (in particular, if $\chi_{0}=0$)
this approach to handling advection fails and one must introduce some
form of \emph{artificial dissipation} in the discrete advection procedure.
In the future we will explore more sophisticated minimally-dissipative
semi-Lagrangian advection schemes\textbf{ }\citet{SemiLagrangianAdvection_2D,SemiLagrangianAdvection_3D}
to handle the case of no bare diffusion.

The filtering of the discrete random velocity field $\V v^{n}$ required
to generate $\V u^{n}$ can be done in one of several ways. The first
approach, which we have employed in several prior works on fluctuating
hydrodynamics \citet{LLNS_S_k,DiffusionRenormalization,LLNS_Staggered,LowMachExplicit},
is to not perform any filtering. This approach was used when preparing
Figs. \ref{fig:DiffusiveInterface} and \ref{fig:SmoothDecay} and
\ref{fig:S_k_c}. In this case the filtering comes from the truncation
of the fluctuating fields on the scale of the grid, that is, $\sigma\approx\D x$,
where $\D x$ is the grid spacing. In this case, it is possible to
explicitly compute the diffusion enhancement for the spatially-discretized
equations by a discrete analog of (\ref{eq:chi_r_Stokes}). This tedious
technical calculation will not be presented here for brevity, and
we only quote the result in two dimensions. We obtain that the effective
diffusion coefficient for the average concentration in the discrete
setting is
\begin{equation}
\chi_{\text{eff}}^{2D}\approx\chi_{0}+\frac{k_{B}T}{4\pi\eta}\ln\frac{L}{\alpha\D x},\label{eq:discrete_2D}
\end{equation}
where $L\gg\D x$ is the length of the square periodic cell (note
that for non-square unit cells the diffusion enhancement is not isotropic).
Here the coefficient $\alpha=1.2$ was estimated by computing the
inverse of the discrete Stokes operator numerically. The formula (\ref{eq:discrete_2D})
is the discrete equivalent of (\ref{eq:chi_SE}). In three dimensions,
\begin{equation}
\chi_{\text{eff}}^{3D}\approx\chi_{0}+\frac{k_{B}T}{\eta\alpha\D x},\label{eq:discrete_3D}
\end{equation}
where the coefficient $\alpha$ can be obtained numerically.

An alternative approach to filtering of the velocity was used when
preparing Fig. \ref{fig:LagrangianEulerian}. Namely, in order to
directly compare to the Lagrangian algorithm described next, we convolved
the discrete advection velocity with an isotropic Gaussian filter
using a multiplication in Fourier space. In this case the width (standard
deviation) of the Gaussian filter $\sigma$ needs to be substantially
larger than the grid spacing $\D x$, for example, $\sigma\gtrsim6\D x$,
in order to obtain a discrete velocity field that is smooth on the
scale of the grid. In such over-resolved simulations one is essentially
solving the continuum equations to a very good approximation. Note,
however, that the resulting algorithm is not efficient because of
the large grid sizes required to resolve the continuum fields with
the grid. In practice, if additional filtering of the discrete velocity
field is desired, it is much more efficient to perform local partial
filtering of the random velocity field using local averaging over
two or three neighboring grid cells, as described in the Appendix
of Ref. \citet{LowMachExplicit}. Such filtering would change the
coefficient $\alpha$ in (\ref{eq:discrete_2D},\ref{eq:discrete_3D})
but not affect the form of the discrete Stokes-Einstein relation.

In order to provide a rough estimate of the relative efficiency of
simulating the overdamped dynamics (\ref{eq:limiting_Strato}) versus
simulating the original dynamics (\ref{eq:v_eq},\ref{eq:c_eq_original}),
let us compare the typical time step size for the Eulerian algorithm
developed above with the method developed in Ref. \citet{LLNS_Staggered}.
Both methods use an FFT-based implicit fluid solver, which dominates
the computational cost, so the cost per time step is similar in the
two algorithms. Resolving the inertial fluid dynamics requires a time
step on the order of $\D t_{\nu}\sim\D x^{2}/\nu$, the time it takes
for momentum to diffuse across one grid cell. The algorithm summarized
above is limited in time step size primarily by the advective Courant
number $v\D t/\D x<1$, where $v$ is a typical magnitude of the advective
velocity obtained by solving the steady Stokes equations with random
forcing. We can estimate $v$ by noting that it is the fictitious
or effective velocity of a particle diffusing with coefficient $\chi$,
$v\sim\sqrt{\chi\D t}/\D t\sim\sqrt{\chi/\D t}$. This gives a time
step limit $\D t\sim\left(\D x^{2}/\chi\right)=\text{Sc}\,\D t_{\nu}$,
which is nothing more than the time it takes a tracer to diffuse across
a grid cell. This shows that simulating the limiting dynamics is $O\left(\text{Sc}\right)$
times faster than the original dynamics.

\subsection{\label{sub:LagrangianAlgorithm}Lagrangian Algorithm}

In the absence of bare dissipation, a faithful discretization of the
overdamped equations must resort to a Lagrangian discretization of
advection. Here we present an algorithm that solves the limiting Lagrangian
equation (\ref{eq:Lagrangian_Strato}) with all truncation errors
strictly controlled to be within numerical roundoff (more precisely,
to twelve decimal places when using double-precision arithmetic) and
\emph{without} artificial dissipation. Such high numerical accuracy
is possible by using a spectral representation of the random flow
and the non-uniform fast Fourier transform \citet{NUFFT}. Note however
that the Lagrangian algorithm is limited in efficiency by the number
of tracers required to represent the finest scales, which grows with
time, as shown in Fig. \ref{fig:LagrangianInterface}. For finite
collection of tracers, e.g., a finite number of colloids in a periodic
box, the Lagrangian algorithm below can be seen as an alternative
to more standard Brownian/Stokesian Dynamics that naturally accounts
for the effects of confinement \citet{BrownianBlobs}. Notably, the
algorithm below scales perfectly \emph{linearly} in the number of
particles. This has enabled us to do simulations with several million
particles. Note that we do not include the effect of gravity (buoyancy)
in the Lagrangian algorithm.

The random velocity is smoothed by convolution with an isotropic Gaussian
filter $\M{\sigma}$, which can be performed as multiplication in
Fourier space. The tracer Lagrangian dynamics (\ref{eq:Lagrangian_Strato})
can be efficiently simulated using the following Lagrangian algorithm
with time step size $\D t$, which updates the tracer positions from
time step $n$ to time step $n+1$ (denoted here by superscript):
\begin{enumerate}
\item Generate a random advection velocity by solving the steady Stokes
equations with random forcing in the Fourier domain
\begin{align*}
i\V k\hat{\pi}_{\V k}^{n+\frac{1}{2}} & =-\eta k^{2}\hat{\V v}_{\V k}^{n}-\sqrt{\frac{2\eta\, k_{B}T}{\D t}}i\V k\cdot\widehat{\M W}_{\V k}^{n}\\
\V k\cdot\hat{\V v}_{\V k}^{n} & =0,
\end{align*}
using a grid of $N^{d}$ wave-indices $\V k$ consistent with the
periodicity. Note that different wave-indices decouple in the Fourier
basis and the above procedure requires only solving a linear system
of $d$ equations for every wavenumber.
\item Filter the velocity with a Gaussian filter (in Fourier space),
\[
\hat{\V w}_{\V k}^{n}=\hat{\M{\sigma}}_{\V k}\hat{\V v}_{\V k}^{n}.
\]
Note the Fourier transform $\hat{\M{\sigma}}$ of a Gaussian filter
$\M{\sigma}$ with standard deviation $\sigma$ is also a Gaussian
with standard deviation $\sigma^{-1}$.
\item Use the non-uniform FFT \citet{NUFFT} to evaluate the velocity at
the locations of the tracers, $\V u^{n}=\V w^{n}\left(\V q^{n}\right)$.
\item Move the tracers using a forward Euler-Maruyama step,
\[
\tilde{\V q}^{n+1}=\V q^{n}+\V u^{n}\D t+\sqrt{2\chi_{0}\D t}\,\V W_{\V q}^{n},
\]
where $\V W_{\V q}^{n}$ are i.i.d. standard Gaussian variates generated
independently for each particle at each time step.
\item For periodic domains there is no difference between different stochastic
interpretations of the Lagrangian equations, and one can set $\V q^{n+1}=\tilde{\V q}^{n+1}$.
For non-periodic domains, one has to perform a corrector step,
\[
\V q^{n+1}=\V q^{n}+\left(\V u^{n}+\tilde{\V u}^{n+1}\right)\frac{\D t}{2}+\sqrt{2\chi_{0}\D t}\,\V W_{\V q}^{n},
\]
where $\tilde{\V u}^{n+1}=\V w^{n}\left(\tilde{\V q}^{n+1}\right)$.
\end{enumerate}
The key to obtaining near roundoff accuracy is the choice of the number
of Fourier modes used to represent the fluctuating velocity field.
Assume that the Gaussian filter $\hat{\M{\sigma}}$ decays to roundoff
tolerance above a wavenumber $k_{0}\approx3\sigma.$ This means that
the Stokes equations only need to be solved for wavenumbers smaller
than $k_{0}$. In order to also be able to perform the non-uniform
FFT with twelve digits of accuracy using a uniform FFT as in the algorithm
we use \citet{NUFFT}, it is necessary to include redundant modes
and set the cutoff wavenumber to $2k_{0}$. This determines the size
of the grid used to perform the forward and inverse FFT transforms
to $N>2k_{0}L/\pi,$ which can be a large number but the algorithm
is easily parallelized on GPUs.

For a sufficiently small time step $\D t\ll\sigma^{2}/\chi_{\text{eff}}$,
the Lagrangian algorithm described here solves the continuum equations
to high accuracy, and the diffusion enhancement can be determined
from the continuum formula (\ref{eq:chi_r_Stokes}) in Fourier space.
For a Gaussian filter of standard deviation $\sigma$ in two dimensions
the relation (\ref{eq:discrete_2D}) with $\D x=\sigma$ holds, where
we numerically estimate the coefficient $\alpha\approx5.5$.


\end{document}